 \documentclass[11pt,twoside]{article}
 \usepackage{atmp}
       
\renewcommand{\copyrightnotice}[4]{
  \leftline{\copyright~ #1 by the authors}
 \leftline{Adv.\ Theor.\ Math.\ Phys. {\bf #2} (#1) #3--#4}
 \setcounter{gfirstpage}{#3}
 \setcounter{glastpage}{#4}
}
\copyrightnotice{2003}{7}{667}{710}

\theoremstyle{plain}

\newtheorem{thm}{THEOREM}[section]
\newtheorem{lm}[thm]{LEMMA}

\theoremstyle{definition}

\theoremstyle{remark}
\newcommand{\upchi}{\raise1pt\hbox{$\chi$}}
\newcommand{\R}{{\mathord{\mathbb R}}}

\newcommand{\C}{{\mathord{\mathbb C}}}

\newcommand{\F}{{\mathcal{F}}}
\newcommand{\J}{{\mathcal{J}}}

\numberwithin{equation}{section}

\begin{document}
\setcounter{page}{667}
\title{Existence of Atoms and Molecules in Non-Relativistic Quantum
Electrodynamics}

\url{math-ph/0307046}

\author{ Elliott H. Lieb$^1$ and Michael Loss$^2$}
\address{$^1$Departments of Physics and Mathematics, Jadwin
Hall\\
Princeton University, P.~O.~Box 708, Princeton, NJ
  08544}
\addressemail{lieb@princeton.edu}

\address{$^2$School of Mathematics, Georgia Tech,
Atlanta, GA 30332}
\addressemail{loss@math.gatech.edu}

\markboth{\it EXISTENCE OF ATOMS  IN NON-RELATIVISTIC QED}{\it E.H. LIEB AND
M. LOSS}


\footnotetext
[1]{Work partially
supported by U.S. National Science Foundation
grant PHY 01-39984.}
\footnotetext
[2]{Work partially
supported by U.S. National Science Foundation
grant DMS 03-00349.\\
\mbox{\ \ \ \ \ \ }This paper may be reproduced, in its
entirety, for non-commercial purposes.}

\begin{abstract}
We show that the Hamiltonian describing $N$ nonrelativistic electrons
with spin, interacting with the quantized radiation field and several
fixed nuclei with total charge $Z$, has a ground state when $N
<Z+1$.  The result holds for any value of the fine structure constant
$\alpha$ and for any value of the ultraviolet cutoff $\Lambda$ on the
radiation field. There is no infrared cutoff. The basic mathematical
ingredient in our proof is a novel localization of  the
electromagnetic field in such a way that the errors in the energy are
of smaller order than $1/L$, where $L$ is the localization radius.

\end{abstract}

\cutpage

\section{Introduction} \label{intro}

The existence of atoms and molecules in the framework of the
Schr\"odinger equation was proved by Zhislin \cite{ZG} for fixed
nuclei when $N <Z+1$. That is to say, the bottom of the spectrum of
the $N$-electron Hamiltonian is a genuine $N$-particle bound state
that satisfies Schr\"odinger's equation with some energy $E$, for each
choice of the locations of the nuclei. (Here $N$ is the number of
electrons, each of charge $-e$, and $Ze$ is the total charge of one or
more fixed, positively charged nuclei.)  The main physical result of
the present paper is the proof of the same thing when account is taken
of the ever-present quantized electromagnetic field. The interaction
of this field with the electrons (but not the field itself)
necessarily has an ultraviolet cutoff $|k|\leq \Lambda$ (in order to
have finite quantities), but we emphasize that no infrared cutoff is
used here.

If the fine structure constant $\alpha = e^2/\hbar c$ and $\Lambda$
are small enough, the result follows from \cite{BFS}, but our result
holds for {\it all values} of these parameters. In a recent paper
Barbaroux, Chen and Vugalter \cite{babar} developed a new method that
shows the existence of ground states for two-electron molecules with
$2<Z+1$ (e.g., the Helium atom). Although they do not have to require
that the perturbation is small when compared to the ionization energy
as in \cite{BFS}, they have to impose restrictions on the various
parameters since their works relies on the existence of the zero
momentum ground state of the Hamiltonian of an electron interacting
only with the radiation field.  This has been established in
\cite{chen} but only for sufficiently small coupling constants.  The
method of \cite{babar} is different from ours.

Our work (and \cite{babar}) relies on earlier work with Griesemer
\cite{GLL} where it was shown that a ground state exists provided a
``binding condition" is satisfied, and it is this condition that is
proved in \cite{babar} for the restricted $N=2$ case and for the
general case here for $N<Z+1$.  If $E^V(N)$ denotes the bottom of the
spectrum of the Hamiltonian $H^V(N)$, which includes the Coulomb
attraction of the electrons to the fixed nuclei of various positive
charges $Z_1e,\ \dots,\  Z_Ke$ with $Z=\sum Z_j$, and if $E^0(N)$ denotes
the bottom of the spectrum of $H^0(N)$ --- the ``free-electron"
Hamiltonian in which there are no nuclei, but the electron-electron
Coulomb repulsion is included --- then the {\it binding condition} is
\begin{equation} \label{binding}
\boxed{
E^V(N) < \min \left\{ E^V(N') + E^0(N-N') : 0\leq N'<N \right\}\ . }
\end{equation}
This binding condition, incidentally, is the same condition that
Zhislin derived for the Schr\"odinger equation without the
quantized electromagnetic field, and which he verified for $N<Z+1$.

The inclusion of the quantized electromagnetic field presents two main
difficulties.  One is that if the bottom of the spectrum contains an
eigenvalue it is {\it not} an isolated eigenvalue, as it was in
\cite{ZG}.  Rather, the bottom of the spectrum is always the bottom of
the essential spectrum because one can create arbitrarily many,
arbitrarily soft photons.  It is not easy to find an eigenvalue when
it lies in the continuum.  This problem was solved in \cite{GLL} under
condition (\ref{binding}).

The second main problem, which complicates the proof of
(\ref{binding}), comes from the fact that each electron carries a
virtual cloud of photons.  This cloud may have substantial energy and
when two electrons are near each other (whether bound or not) the
interference of the photon clouds must be taken into account.  In
general, this is a highly non-perturbative effect.  Our way around
this difficulty is to prove that the photon clouds can be localized
(i.e., effectively eliminated outside a ball of radius $L$ surrounding
the electron or the atom) in such a way that the error induced in the
energy of the cloud is smaller than $L^{-(1+\varepsilon)}$, and thus
the direct Coulomb interaction, which goes as $L^{-1}$, is dominant
--- as it was in the original paper \cite{ZG}.  A closely related
effect is that even in the absence of an external potential electrons
interact with each other.  In such a case their dynamics is governed
by $H^0$, which contains the electron-electron Coulomb repulsion.
Nevertheless, it is not inconceivable that the quantized field, which
interacts simultaneously with all the electrons, might cause binding
among the ``free'' electrons.  While this is unlikely it has never
been disproved and we must not assume in our proof that $E^0(N) =
NE^0(1)$.

We are grateful to the anonymous referee who made many valuable
suggestions and who helped us understand some conceptual matters about
photons.


\section{Basic Definitions and Concepts} \label{main}

The Hamiltonian under consideration, in appropriate units, is the
Pauli-Fierz
Hamiltonian and is given by
\begin{align}
H^V(N) &= \sum_{j=1}^N \left[ (p_j + \sqrt{\alpha} A(x_j))^2
+\frac{g}{2}
\sqrt{\alpha}\> \sigma_j \cdot B(x_j) + V(x_j) \right] \nonumber \\
&\qquad\qquad +\alpha\sum_{i < j}\frac {1}{|x_i-x_j|} + H_f \ .
\end{align}
Here, $g$ is some constant (close to 2, physically) and the vector
$\sigma_j$ is the set of three Pauli spin matrices for electron $j$.
(Owing to the ultraviolet cutoff there is no restriction to $|g|\leq
2$, as there would be without a cutoff \cite{FLL}.) The operator $p_j$
denotes $-i\nabla$ acting on the coordinate of the $j$-th electron.

The potential $V$ is the potential of $K\geq 1$ nuclei with positive charges
$Z_1,\dots, Z_K $ and locations $R_1,\dots, R_K \in {\mathbb R}^3$.
\begin{equation}\label{vee}
V(x) = -\sum_{j=1}^K Z_j |x-R_j|^{-1}\ .
\end{equation}

{\it Remark:} The truth of our main theorem (\ref{atom}) --- and its
proof --- does not require that $V(x)$ be given by (\ref{vee}). In
addition to the general condition \cite[eq. (5)]{GLL}, we need only the
condition that there is some radius $\rho$ such that $\langle V(x)
\rangle \leq -Z/|x|$ for $|x| > \rho$, where $\langle \ \cdot \
\rangle$ denotes spherical average. Similarly, the repulsion
$|x_i-x_j|^{-1}$ can be replaced by $W(x_i -x_j)$ provided $\langle
W(x) \rangle \leq 1/|x|$ for all $|x|>\rho$.

The free Hamiltonian $H^0(N)$ is similar to $H^V(N)$, but without the
attraction to the nuclei, i.e.,
\begin{equation}\label{free}
H^0(N) = \sum_{i=1}^N \left[(p_i + \sqrt{\alpha} A(x_i))^2
+\frac{g}{2} \sqrt{\alpha}\> \sigma_i \cdot B(x_i)
\right] + \alpha\sum_{i < j} \frac{1}{|x_i-x_j|} + H_f \ .
\end{equation}
Note that the Coulomb repulsion among the electrons is included. The
reason for including the Coulomb repulsion is (as stated above) that
we do not know whether the electrons bind to each other through the
interaction with the electromagnetic field, i.e., the electrons may
not separate in the lowest energy state (if there is one).

The (ultraviolet cutoff) magnetic vector potential
is defined by
\begin{equation}
A(x) = \frac{1}{2\pi} \sum_{\lambda=1}^2 \int
\frac{\varepsilon_\lambda(k)}{\sqrt{|k|}}
\widehat\chi_{\Lambda}^{\phantom{B}}(k) \left(
\widehat a^{{\phantom{\ast}}}_\lambda(k) e^{ik\cdot x}
+\widehat a_\lambda^{\ast}(k) e^{-ik\cdot x}\right)dk
\  ,
\label{apot}
\end{equation}
where the function $\widehat\chi_{\Lambda}^{\phantom{B}}$ is a smooth,
radial function in $k$ space, that vanishes outside the ball whose
radius is the ultraviolet cutoff $\Lambda$. We denoted the creation
and destruction operators of photons of momentum $k$ and polarization
$\lambda$ by $\widehat a_\lambda (k)$ and $\widehat a_\lambda^* (k)$.
This unusual notation is used since we shall later introduce the
creation and destruction operators in configuration space, $ a_\lambda
(y)$ and $a_\lambda^* (y)$, which act on the Fourier transformed
functions in Fock space.

The magnetic field is $B(x) = {\mathrm{curl}}A(x)$. The operators
$\widehat a^{{\phantom{\ast}}}_{\lambda}, \widehat a^{\ast}_{\lambda}$
satisfy the usual commutation relations
\begin{equation}
[\widehat a_\lambda(k), \widehat a^{\ast}_{\nu} (q)] = \delta ( k-q)
\delta_{\lambda, \nu}\ , ~~~ [\widehat a^{{\phantom{\ast}}}_
{\lambda}(k), \widehat a^{{\phantom{\ast}}}
_{\nu} (q)] = 0, \quad {\rm{etc}}
\end{equation}
and the vectors $\varepsilon_{\lambda}(k)$ are the two possible
orthonormal polarization vectors perpendicular to $k$ and to each
other.

The vectors $\varepsilon_\lambda(k)$
have to be discontinuous functions of k on every sphere of fixed
$|k|$-value because it is not possible to
``comb the hair on a sphere". However, the only physical quantity,
\begin{equation}
\sum_{\lambda = 1}^2 \varepsilon^i_{\lambda}(k) \varepsilon^j_{\lambda}(k)
= \delta_{i,j} - \frac{k_i k_j}{|k|^2}\  ,
\end{equation}
is discontinuous only at the point $k=0$. For the rest of this paper
we choose the polarizations vectors to be
\begin{eqnarray}\label{polvec}
\varepsilon_1(k) &=& \frac{(k_2, -k_1, 0)}{\sqrt{k_1^2+k_2^2}} \ ,\nonumber
\\
\varepsilon_2(k) &=& \frac{k}{|k|} \wedge  \varepsilon_1(k) \ .
\end{eqnarray}

Let us emphasize here that some smoothness of the function
$\widehat\chi_{\Lambda}^{\phantom{B}}$ is essential for our arguments
since this guarantees that the  coupling functions
\begin{equation}\label{couplingfunction}
h^i_\lambda (y)=\frac{1}{2\pi} \int
\frac{\widehat\chi_{\Lambda}^{\phantom{B}}(k)} {\sqrt{|k|}}
\varepsilon^i_\lambda (k) e^{-ik
\cdot y} dk
\end{equation}
has a suitable decay as $|y| \to \infty$. If we did not have the
discontinuous function $\varepsilon^i_\lambda (k)$ in
(\ref{couplingfunction}) then $h(y)$ would decay as $|y|^{-5/2}$ as
$|y| \to \infty$.  (Proof: $|k|^{-1/2}$ is the Fourier transform of
$|y|^{-5/2}$ in the sense of distributions \cite[Theorem
5.9]{Analysis}. The Fourier transform of $\widehat\chi_{\Lambda}$ is
real analytic and decays faster than any inverse power of $|y|$.
Hence, the convolution of $\chi$ with $|y|^{-5/2}$ decays like
$|y|^{-5/2}$.  With a sharp cutoff it would decay only like $|y|^{-2}$
which turns out to be insufficient for a good localization of the
photon states.

This analysis of $h$ shows that we have to be circumspect about the
choice of the polarization vectors.  Their discontinuity will spoil
the $|y|^{-5/2}$ decay, but it is important to get better decay than
$|y|^{-2}$.  In Lemma \ref{coupling} of Appendix B it is shown that
with our choice (\ref{polvec}) of the polarization vectors the
coupling functions have sufficient decay in the sense that $\int
|y|^{2\gamma} |h^i_\lambda(y)|^2 dy $ is finite for all $\gamma <1$.
Thus, in an average sense, the coupling functions decay almost as fast
as $|y|^{-5/2}$.  We made no attempt to optimize the choice of the
polarization vectors.

While polarization is physically measurable, the polarization vectors
are not.  They are merely a basis.  It is odd, therefore, that their
mathematical definition plays a role in the spatial localization of
the photon field that we shall construct, and which is central to our
proof of the binding condition.  It would be better to start with a
formalism that contains only ``divergence-free" vector fields as the
dynamical variables instead of trying to define them with the aid of
unphysical polarization vectors.  In particular, the Fock space would
be built over the $L^2$-space of divergence-free vector fields instead
of $L^2\otimes \mathbb{C}^2$.  We shall not explore this here, but we
mention that the localization of a divergence-free vector field, which
preserves the divergence-free property is also a subtle matter.

The field energy, $H_f$, sometimes called $d\Gamma(\omega)$, is given by
\begin{equation}\label{eq:fielden}
H_f = \sum_{\lambda=1,2} ~ \int_{\R^3} ~ |k|
\widehat a^*_\lambda(k) \widehat a_\lambda (k) d k
\end{equation}
There is no cutoff in $H_f$. The energy of a photon is $|k|$.

Another unbounded operator of interest is the number operator
\begin{equation}\label{numop}
\mathcal{N} =  \sum_{\lambda=1,2} ~ \int_{\R^3} ~
\widehat a^*_\lambda(k) \widehat a_\lambda(k) d k \ .
\end{equation}

The physical Hilbert space for this system is given by
\begin{equation}
\mathcal{H}(N) = \wedge^N L^2(\mathbb{R}^3; \mathbb{C}^2) \otimes
\mathcal{F}
\end{equation}
where the wedge indicates that the electron wave functions are
antisymmetric under the exchange of the particle labels. Thus, the
functions in the space $\mathcal{H}(N)$ obey the Pauli exclusion
principle. The photon Fock space is $\mathcal{F}$.
We denote the inner product of two states $\Psi$ and $\Phi$
in the space $\mathcal{H}(N)$  or in Fock space alone by
\begin{equation}
\left(\Psi, \Phi \right) \quad\quad {\mathrm{and}}\quad\quad
 \left\langle \Psi, \Phi \right\rangle \  ,
\end{equation}
respectively. If $\Psi$ and $\Phi$ are in $\mathcal{H}(N)$ then
$\left\langle \Psi, \Phi \right\rangle$  makes sense and defines
a summable function of $x_1, s_1, \ \dots \ , x_N, s_N$, where
$x_j, s_j$ are the space-spin variables of the $j$-th electron.

It is desirable that the above Hamiltonians be selfadjoint on certain
domains and this has been worked out, e.g., in \cite{Hiro}. In this
paper we will always be talking about the Friedrichs extension of the
symmetric operators $H^\#(N)$ (where $\#$ is $0$ or $V$). The form
domain will consist of all states for which each term in the operators
has a finite expectation value.  Accordingly, we define the ground
state energy $E^\#(N)$ for the Hamiltonian $H^{\#}(N)$ by
\begin{equation}
E^\#(N) = \inf\left\{\left(\Psi, H^\#(N) \Psi \right) :
\Psi \in \mathcal{H}(N) \ ,
\Vert \Psi \Vert = 1 \right\} \  .
\end{equation}
The numbers $E^\#(N)$ are  finite. This follows from Lemma A.4 in \cite{GLL}
together with the fact that
the Coulomb potential is form bounded with respect to $p^2=-\Delta$.

A few remarks concerning the Fock space $\F$ are in order.
It is built over the space
$L^2(\mathbb{R}^3) \otimes \mathbb{C}^2$; the second factor takes
into account the polarizations.
Let $\{f_i\}$, $i = 1,2,\dots$, be an orthonormal basis for
$L^2(\mathbb{R}^3) \otimes \mathbb{C}^2$. Then, vectors of the form
\begin{equation} \label{basis}
|i_1,m_{1}; \ \dots \ ; i_n, m_{n} \rangle =\frac{1}{\sqrt{m_1! \cdots
m_n!}}
\ a^*(f_{i_1})^{m_{1}} \cdots a^*(f_{i_n})^{m_{n}} |0\rangle \ ,
\end{equation}
constitute an orthonormal basis for $\F$, the occupation number basis.  In
(\ref{basis}) $n$ is an arbitrary nonnegative integer (with $n=0$
denoting the vacuum vector $ |0\rangle$) , the indices $i_1, \cdots,
i_n$ are all different, the $m_i$ are all positive integers,
$a^*(f) $ is an abbreviation for
$\sum_{\lambda}a^{{{\ast}}}_{\lambda}(f_{\lambda})$ and
$f_\lambda=f(k,\lambda)$
is a function in $L^2$.  Thus, any state
$\Phi \in \F$ can be uniquely written as
\begin{equation}\label{rep}
\Phi = \sum_{n\geq 0}\  \sum_{i_1 < i_2 < \ \cdots \ < i_n} \
\sum_{m_{1},  \ \dots \ , m_{n} }
\phi_{i_1,m_1;\  \dots \  ; i_n,m_n} \
|i_1,m_{1}; \ \dots \ ; i_n,m_{n} \rangle \ ,
\end{equation}
where the $n=0$ term in (\ref{rep}) is just $\phi_0 |0\rangle$ with
$\phi_0 \in \mathbb{C}$. The inner product is given by
\begin{equation}
\left\langle \Phi, \Phi \right\rangle =\sum_{n\geq 0}\
\sum _{i_1 < i_2 < \ \cdots \ <i_n} \
\sum_{m_{i_1},
\ \dots \ , m_{i_n} }\ |\phi_{i_1,m_1; \ \dots \ ; i_n,m_n}|^2 \ .
\end{equation}

This representation has the advantage that the symmetry in the photon
variables is automatically taken care of. It is particularly useful
when dealing with product states. Consider a state $\Phi$ whose
photons are all {\bf localized} in a closed region $\mathcal{Y}\subset
\mathbb{R}^3$. This means that all the $f_i(y,\lambda)$ appearing in
(\ref{basis}) and in (\ref{rep}) vanish if $y \notin \mathcal{Y}$.
Likewise, consider a state $\Psi$ whose photons are all localized in a
closed region $\mathcal{Z}\subset \mathbb{R}^3$ which is disjoint from
$\mathcal{Y}$. Pick an orthonormal basis $\{f_k\}$ in
$L^2(\mathcal{Y}) \otimes \mathbb{C}^2$ and an orthonormal basis
$\{g_\ell\}$ in $L^2(\mathcal{Z}) \otimes \mathbb{C}^2$. Clearly, the
two algebras of creation and annihilation operators generated by
$a^{\#}(f_k)$ and $a^{\#}(g_\ell)$ commute. If
\begin{equation}
\Phi = \sum_{n\geq 0}\  \sum_{i_1 < i_2 < \ \cdots \ < i_n} \
\sum_{p_{1}, \ \dots \ ,  p_{n} }\ \phi_{i_1,p_1; \
\dots \ ; i_n,p_n} \ \ |i_1,p_{1};
\ \dots \ ; i_n,p_n
\rangle^{\phantom{x}}_{\mathcal{Y}}
\end{equation}
and
\begin{equation}
\Psi = \sum_{n\geq 0}\  \sum_{j_1 < j_2 < \ \cdots\  < j_n} \
\sum_{q_{1},  \ \dots \ ,  q_{n} }\ \psi_{j_1,q_1; \
\cdots \  ; j_k,q_k} \ \  |j_1,q_{1};\  \dots \ ; j_k,q_k
\rangle^{\phantom{x}}_{\mathcal{Z}}
\end{equation}
then we define the product state $\Xi$ by
\begin{multline} \label{product}
\Xi = \\
\sum \phi_{i_1,p_1; \ \cdots \  ; i_m,p_m}
\psi_{j_1,q_1; \ \cdots \ ; j_k,q_k}
\ \ |i_1,p_{1};\ .\ . \ ; i_m,p_{m}
\rangle^{\phantom{x}}_{\mathcal{Y}}\otimes
\ |j_1,q_{1}; \ .\ . \ ; j_k,q_{k} \rangle^{\phantom{x}}_{\mathcal{Z}}
\end{multline}
where
\begin{multline}
 |i_1,p_{1}; \ \cdots \ ; i_m,p_{m} \rangle^{\phantom{x}}
_{\mathcal{Y}}\otimes |j_1,q_{1};  \ \cdots \ ; j_k,q_{k}
\rangle^{\phantom{x}}_{\mathcal{Z}} =  \\
\frac{1}{\sqrt{p_1! \cdots p_m!}\sqrt{q_1! \cdots q_k!}} \,
a^*(f_{i_1})^{p_{1}}
\cdots a^*(f_{i_m})^{p_{m}}
a^*(g_{j_1})^{q_{1}} \cdots a^*(g_{j_k})^{q_{k}} |0\rangle
\end{multline}

By a simple calculation
we find that
\begin{equation}
\left\langle \Xi, \Xi \right\rangle = \left\langle \Phi, \Phi \right\rangle
 \left\langle \Psi, \Psi \right\rangle \ .
\end{equation}
Further, if $f$ is a function supported in $\mathcal{Y}$, then
\begin{equation}
\left\langle\Xi, a^*(f) a(f)\Xi \right\rangle = \left\langle\Phi, a^*(f)
a(f)\Phi
\right\rangle
\left\langle \Psi, \Psi \right\rangle \ .
\end{equation}
Likewise, if $f$ is supported in $\mathcal{Y}$ and $g$ in $\mathcal{Z}$,
then
\begin{equation}
\left\langle\Xi, a^*(f) a(g)\Xi \right\rangle = \langle \Phi, a^*(f)\Phi
\rangle
\left\langle \Psi, a(g)\Psi \right\rangle \ .
\end{equation}

Quite generally, we have, for normal-ordered, bilinear expressions,
the following formulas (in which $\beta$, $\gamma$ denote linear forms
in the annihilation operators $a$, and hence $\beta^*$, $\gamma^*$ are
linear forms in the creation operators):

\begin{align} \label{bg}
\left\langle \Xi, \beta\, \gamma\, \Xi\right\rangle
&=\left\langle \Psi, \beta\, \gamma\, \Psi\right\rangle
\left\langle \Phi, \Phi \right\rangle
+ \left\langle \Psi, \Psi \right\rangle \left\langle
\Phi, \beta\, \gamma\, \Phi\right\rangle \nonumber \\
&\quad +\left\langle \Psi, \beta\, \Psi \right\rangle \left\langle
\Phi, \gamma\,\Phi\right\rangle
+ \left\langle \Psi, \gamma\, \Psi \right\rangle \left\langle \Phi,
\beta\, \Phi \right\rangle  \nonumber    \\
\left\langle \Xi, \beta^*\, \gamma^*\, \Xi \right\rangle
&= \left\langle \Psi, \beta^*\, \gamma^* \, \Psi \right\rangle
\left\langle \Phi, \Phi \right\rangle
+ \left\langle \Psi, \Psi \right\rangle \left\langle \Phi, \beta^*\,
\gamma^*\, \Phi\right\rangle \nonumber \\
&\quad +\left\langle \Psi, \beta^*\, \Psi \right\rangle \left\langle
\Phi, \gamma^*\,
\Phi \right\rangle
+ \left\langle \Psi, \gamma^*\, \Psi \right\rangle
\left\langle \Phi, \beta^*\, \Phi \right\rangle      \nonumber  \\
\left\langle \Xi, \beta^*\, \gamma\, \Xi \right\rangle
&= \left\langle \Psi, \beta^*\,
\gamma\, \Psi\right\rangle
\left\langle \Phi, \Phi \right\rangle + \left\langle \Psi, \Psi
\right\rangle
\left\langle \Phi,
\beta^*\, \gamma\, \Phi\right\rangle  \nonumber \\
&\quad +\left\langle \Psi, \beta^*\, \Psi \right\rangle \left
\langle \Phi, \gamma \,
\Phi\right\rangle +
\left\langle \Psi, \gamma \, \Psi \right\rangle
\left\langle \Phi, \beta^*\, \Phi \right\rangle
\end{align}

A formula of the type (\ref{bg}) does {\it not exist} for
anti-normal-ordered products $\beta\, \gamma^*$. We shall have no need
of such terms, however, because the only source of such terms is
$A(x_i)^2$ in the electron kinetic energy. If we denote the part of
(\ref{apot}) coming from $\widehat a_\lambda(k)$ by $\beta(x)$ and the
remainder by $\beta^*(x)$ (see (\ref{adecomp})) then
\begin{equation}
A(x)^2 = \beta(x)^2+ \beta^*(x)^2 + 2\beta^*(x)  \beta(x) + C
\end{equation}
where
\begin{equation} \label{cee}
C= \frac{1}{2\pi^2} \int
\frac{|\widehat{\chi}_\Lambda^{\phantom{B}}(k)|^2} {|k|} dk \  .
\end{equation}
Thus, apart from a fixed, finite number $\alpha NC$, which is strictly
proportional to $N$, (and which is, therefore, independent of any
decomposition of the system into clusters) we can (and henceforth
shall) replace $A(x_i)^2$ by the normal-ordered

\begin{equation}
:A(x_i)^2:=
\beta(x_i)^2+ \beta^*(x_i)^2 + 2\beta^*(x_i) \beta(x_i)\ .
\end{equation}

Formulas (\ref{bg}) continue to hold for vectors $\Phi$ in the
physical Hilbert space ${\mathcal{H}}(N)$, with the replacement of
$\langle\ , \ \rangle$ by $\left(\ , \ \right)$. In this case, the
coefficients, $ \phi_{i_1,m_1;\  \dots \  ; i_n,m_n} $,
are (antisymmetric) functions of the electron
space-spin coordinates, $x_i, \ s_i$.

It is convenient to introduce the  operators
given by
\begin{equation} \label{avec}
a_\lambda(y)= \frac{1}{(2\pi)^3}\int \widehat a_\lambda(k)
e^{i k \cdot y} d k \ .
\end{equation}
Then the vector potential can be written as
\begin{equation}\label{adecomp}
A^i(x) = \sum_{\lambda =1}^2 a_\lambda (h^i_\lambda (x-\cdot))
+ a_\lambda^{*}(h^i_\lambda(x-\cdot))\ .
\end{equation}

The action of the operators $ a_\lambda (h^i_\lambda(x-\cdot))$ is given by
\begin{multline}
[ a_\lambda (h^i_\lambda (x-\cdot))\Psi]_n(y_1,\lambda_1; \cdots ; y_n,
\lambda_n) \\
 = \sqrt{n+1}\int \overline{h^i_\lambda(x-y)}[\Psi]_{n+1}
(y, \lambda; y_1, \lambda_1;  \cdots ;y_n, \lambda_n) dy \ .
\end{multline}

A convenient expression for $a_\lambda (h^i_\lambda (x-\cdot))$ is the
formula
\begin{equation}
a_\lambda (h^i_\lambda (x-\cdot))= \int a_\lambda(y)
\overline{h^i_\lambda(x-y)} dy \ .
\end{equation}

The number operator and the field energy can be expressed
in terms of the operators $a_\lambda (x)$  by
\begin{equation}
\left(\Phi,\mathcal{N} \Phi\right) = (2\pi)^3 \sum_{\lambda=1}^2
\int \Vert a_\lambda (x) \Phi \Vert^2 dx \ ,
\end{equation}
and
\begin{equation}
\left(\Phi, H_f \Phi\right)= (2\pi)^3\sum_{\lambda =1}^2
\left( a_\lambda(\cdot)\Phi , \sqrt{-\Delta}\  a_\lambda(\cdot)\Phi\right)
\end{equation}
which by eq. 7.12(4) in \cite{Analysis} can be rewritten as
\begin{equation}
4\pi \sum_{\lambda =1}^2 \int \frac{\Vert a_\lambda(x)
\Phi - a_\lambda(y) \Phi \Vert^2 }{|x-y|^4} dx dy \ .
\end{equation}

By the previous considerations we have for the product state $\Xi$
that
\begin{multline}
\sum_\lambda \left(a_\lambda(x)\Xi, a_\lambda(y) \Xi \right) = \\
 \sum_\lambda  \left(a_\lambda(x) \Phi,a_\lambda(y) \Phi \right)
\left(\Psi, \Psi \right)
+\left(\Phi, \Phi \right) \sum_i  \left(a_\lambda(x)
\Psi,a_\lambda(y) \Psi \right ) \\
+\sum_\lambda\left(a_\lambda(x)\Phi, \Phi \right)
\left(\Psi,a_\lambda(y) \Psi \right)
+\sum_\lambda \left(\Phi, a_\lambda(y)\Phi \right)
\left(a_\lambda(x) \Psi, \Psi \right ) \ ,
\end{multline}
and hence we obtain for the field energy of $\Xi$ the expression
\begin{multline} \label{fieldproduct}
\left(\Xi, H_f \Xi \right)  =
\left(\Phi, H_f \Phi \right)\left(\Psi, \Psi \right) +
\left(\Phi, \Phi \right) \left(\Psi, H_f \Psi \right)  \\
-8\pi\sum_\lambda  \Re \left[\int \frac{\left(a_\lambda(x) \Phi,\Phi \right)
\left( \Psi, a_\lambda(y) \Psi \right) +
\left( \Phi,a_\lambda(y)\Phi \right) \left(a_\lambda(x) \Psi, \Psi
\right)}{|x-y|^4}
\right]dx dy \ .
\end{multline}
The $x$ integration in the first term of the last integral runs over
the set $\mathcal{Y}$ while the $y$ integration runs over the set
$\mathcal{Z}$ and similarly in the second term the $x$ integration
runs over the set $\mathcal{Z}$ while the $y$ integration runs over
the set $\mathcal{Y}$.  Hence the last expression is well defined as
long as the distance between the sets $\mathcal{Y}$ and $\mathcal{Z}$
is positive. This term expresses the fact that the field energy is a
nonlocal operator and this nonlocality is one of the main obstacles to
be overcome.

In general, the states $\Phi$ and $\Psi$ will depend on the position
and spin variables of the various electrons and hence the product
state (\ref{product}) has to be antisymmetrized over the electron
labels. It is straightforward to check that the expression
(\ref{fieldproduct}) continues to hold also for such states.  (When
different groups of electrons are involved an antisymmetrization is
required, however, as discussed in (\ref{product2}).)

We need one more concept before stating our main theorem.  It will be
necessary to localize both the electrons and the photon field. As far
as the electrons are concerned it is useful to define what we mean by
a {\bf symmetrized product of $n$ domains} in ${\mathbb R}^3$. If
$B_1,\dots B_n$ are $n$ domains (open sets) in $\R^3$ then the
symmetrized product, $\Omega$, is a domain in $(\R^3)^n$ given by
\begin{equation}
\Omega(B_1,\dots, B_n) =  \bigcup_{\pi \in S_n} B_{\pi 1} \times B_{\pi 2}
\times
\cdots \times B_{\pi n}\  ,
\end{equation}
where $S_n$ is the group of permutations of $n$ labels. It might be
useful to illustrate this when $n=2$. Then we have $B_1 \subset \R^3$,
$B_2 \subset \R^3$ and $\Omega(B_1,B_2) = (B_1 \times B_2) \cup (B_2
\times B_1)$.  This is different from $ (B_1 \cup B_2) \times (B_2
\cup B_1)$. Physically it means there is one particle in each domain
$B_i$, but the label of the particle is indeterminate.  If the domains
overlap there may be several particles in one domain, of course.

\section{The Main Theorem}\label{thm}

The following is our main theorem. The proof given in this section
uses several inequalities derived later on in this paper, but we
present the proof now in order to make the main ideas clear without
too many technicalities.

\begin{thm}[Binding in Atoms]\label{atom}
  The strict inequality (\ref{binding}) holds for all $N<Z+1$, all
  $g$, all $\alpha$ and all $\Lambda$. In particular this implies that
  there exists a normalized ground state $\Phi(N)$ in $\mathcal{H}(N)$
  for the Hamiltonian $H^V(N)$, i.e., $(\Phi(N), H(N) \Phi(N)) =
  E^V(N)$, and it satisfies $H^V(N) \Phi(N) = E^V(N) \Phi(N)$.
\end{thm}

See the remark after eq. (\ref{vee}).

{\it PROOF:} Our proof has three main parts. The first is the
construction of a good trial function for $N-N'$ (with $0 \leq N' < N$)
localized, `free' electrons and localized photons accompanying these
localized electrons. The second part is the construction of a good
trial function for $N'$ localized electrons `bound' to the given,
fixed nuclei, together with localized photons. The third part consists
in the construction of a trial function which is a product of these
two functions and then showing that the energy is lowered (by a
greater amount than the localization errors) because of a negative
Coulomb energy between the `bound' system (consisting of electrons and
nuclei) and the localized `free' electrons. One difficulty in part 3
is that although the photons in the two regions are localized in
separate regions, there is still a residual interaction between the
two fields, given by the last term in (\ref{fieldproduct}), which has to
be considered. This interaction comes from the fact that
multiplication by $|k|$ in Fourier space is a nonlocal operation in
position space.

The general argument proceeds by induction. We know from \cite{GLL}
that one electron binds. Assuming that the binding condition holds for
$M$ electrons, all $1 \leq M \leq N-1$, we have to show that it holds
for $N$ electrons, i.e., $E^V(N) < \min\{E^V(N')+E^0(N-N'): 0 \leq N'
< N\}$.  Using \cite[Theorem 2.1]{GLL} we may assume that the
Hamiltonian $H^V(M)$ has a ground state for all $1 \leq M \leq
N-1$. By the second part of \cite[Theorem 3.1]{GLL} we know that
$E^V(N) < E^0(N)$ for all $N$, since $Z>0$ and the attractive Coulomb
potential is strictly negative.

\boxed{Part 1.} From now on we set
$$
n=N-N'\ .
$$
Given $0 \leq N' <N$ we shall construct a normalized state $\Phi(n)$
for the free electron Hamiltonian $H^0(n)$ with the property that the
$n$ electrons are localized in a symmetrized product $\Omega$ of some
balls of radius $R_0$ while the field is localized in balls with the
same center but with radius $L>2R_0$. The construction of $\Phi(n)$ is
done in Theorem \ref{locphot}. It lies in the physical Hilbert space
${\mathcal H}(n)$ and has an energy given by
\begin{equation}\label{localenergy}
\frac{\left( \Phi(n), H^0(n) \Phi(n) \right)} {\left(
\Phi(n),\Phi(n) \right)}
\leq E^0(n)+\frac{C n}{(L-2R_0)^{\gamma}}\left(\frac{R_0}
{L^\gamma}\right) (1+|\log(\Lambda R_0)|) +
\frac{2\pi^2 n^{\, 2} }{R_0^2}  \ ,
\end{equation}
for any $\gamma <1$, where $C$ is some constant independent of $L$ and
$R_0$. (It does depend on $\gamma$ and on $n$, but $n$ is bounded by
$N$).

In (\ref{localenergy}) the term
$\left(Cn/(L-2R_0)^{\gamma}\right)\left(R_0/L^\gamma \right)
(1+|\log(\Lambda R_0)|) $
comes from the energy needed to localize the field in $n$ balls of
radius $L$.  The last term comes from the kinetic energy needed to
localize $n$ electrons in the $n$ balls of radius $R_0$
(Lemma \ref{locelect}).

\boxed{Part 2.}
According to the induction assumption at the beginning of this proof
we may assume that $1\leq N' \leq N-1$ and that the Hamiltonian
$H^V(N')$ of the bound electrons has a normalized ground state
$\Gamma(N')$.  By \cite[Lemma 6.2]{GLL} we know that this ground state
is exponentially localized in the electron variables, i.e., if we
denote by $|X|$ the quantity $\sum_{i=1}^{N'} |x_i|$ then
\begin{equation}  \label{expdecay}
\Vert e^{\beta |X|} \Gamma(N') \Vert ^2 \leq C_\beta
\end{equation}
for any $\beta^2 <  \min\{E^V(N'-m)+E^0(m): 0 < m \leq N'\} - E^V(N')$.

(Note: An error in the proof of this exponential localization in
\cite[Lemma 6.2]{GLL} was discovered by J-M. Barbaroux and the
necessary correction was published in \cite{G}. We are grateful to
Prof. Barbaroux for pointing out this mistake to us.)

Although it is not necessary to do so, we (strictly) localize
$\Gamma(N')$ so that all the electrons are in a common ball of radius
$R_0$. Following that, we  localize the photon field in a larger
ball of radius $L >R_0$. The field localization is essential. The
electron localization is not since it would be possible to use only
the exponential decay of $\Gamma(N')$. The localization is done as follows.

Let $\chi \leq 1 $ be a smooth cutoff function with support in the unit ball
centered at the origin and $\chi(x) =1$ for $ |x|<1/2 $.  Let
$\Theta = \prod_{i=1}^{N'} \chi (x_i/R_0)$ and
$\widetilde{\Gamma}(N') = \Theta \Gamma(N')$. Since
$\Gamma(N')$ is a ground state, and hence satisfies the Schr\"odinger
equation, we can  deduce that
the increase in energy due to the cutoff is bounded as follows.
\begin{equation} \label{innerenergy}
\left( \widetilde{\Gamma}(N'), H^V(N') \widetilde{\Gamma}(N')\right)
\leq E^V(N') \left( \widetilde{\Gamma}(N'), \widetilde{\Gamma}(N')\right)
+ N' \frac{C_\beta \exp({-\beta R_0})}{R_0^2} \  .
\end{equation}

Inequality (\ref{innerenergy}) follows from (\ref{expdecay}) and
integration by parts as follows.
With $\langle\, , \, \rangle $ denoting inner product in Fock space
and $dX$ denoting integration over the space-spin variables,
we have
\begin{align}\label{intparts}
\int  \langle  \Theta \Gamma(N'),
\sum_j &(\nabla_j +i A(x_j))^2 \Theta \Gamma(N') \rangle \ dX = \nonumber \\
&\quad \quad \int \left(\sum_j \Theta \Delta_{x_j} \Theta \right)
 \langle \Gamma(N'), \Gamma(N')
\rangle \ dX  \nonumber \\
&+2
\int \sum_j \left(\Theta \nabla_{x_j} \Theta\right) \cdot \langle \Gamma(N')
,
(\nabla_j +iA(x_j) )
\Gamma(N') \rangle \ dX \nonumber \\
& \quad +\int \Theta^2 \langle \Gamma(N'),
\sum_j (\nabla_j +i A(x_j))^2
\Gamma(N') \rangle dX \ .
\end{align}
Since $(\nabla +i A(x))^2 $ is a symmetric operator, the left side of
(\ref{intparts}) is real and, therefore, the right side must be real,
too. The first term on the right side is real. The third term is also
real because $\Gamma (N')$ satisfies the Schr\"odinger equation, and
hence $ \langle \Gamma(N'), \sum_j (\nabla_j +i A(x_j))^2 \Gamma(N')
\rangle = \langle \Gamma(N'), (-E^V(N') + {\mathrm{real}\ \mathrm{
    potentials}})\Gamma (N') \rangle$, which is real.  The middle term
must, therefore, be real too (when summed over all particles), and we
can replace the integrand by its real part. This means that we can
replace $2\langle \Gamma(N') , (\nabla_{x_j} +iA(x_j) ) , \Gamma(N')
\rangle$ by $2\Re \langle \Gamma(N') , \nabla_{x_j} \Gamma(N') \rangle
= \nabla_{x_j} \langle \Gamma(N'), \Gamma(N') \rangle$ since
\hfill\break $\langle \Gamma(N') , iA(x_j)\Gamma(N') \rangle$ is
imaginary (because $A(x_j)$ is symmetric).

Now, integrating by parts we can combine the second term with the
first to yield $-\int (\nabla_{x_j} \Theta)^2 \langle \Gamma(N'),
\Gamma (N')\rangle dx_j$.  This is the error term (the last term leads
to the principal term $E^V(N') \left( \widetilde{\Gamma}(N'),
\widetilde{\Gamma}(N')\right) $).  This error term can be bounded by
replacing $\nabla \chi(x_j/R_0) $ by $C/R_0$ times the
characteristic function of the annulus between $R_0/2$ and $R_0$, for
some constant $C$. But this characteristic function is bounded by
$\exp({-\beta R_0/2}) \exp({+\beta |x|})$. Inequality
(\ref{innerenergy}) then follows from the exponential decay
(\ref{expdecay}).

Next one has to show that the error term in (\ref{innerenergy}) is
small when compared with $\Vert \widetilde{\Gamma}(N') \Vert^2$. It
follows from the exponential decay that
\begin{equation}\label{normbound}
\Vert \widetilde{\Gamma}(N') \Vert^2 \geq 1- N'C_\beta e^{-\beta R_0} \ .
\end{equation}
To see this note that $\chi(x_i/R_0)^2 \geq 1-g_i$, where
$g_i=1$ if $|x_i| >R_0/2$ and $g_i=0$ otherwise. Then
$\Theta \geq \prod_i (1-g_i) \geq 1-\sum_i g_i$. But
$g_i \leq \exp\{-\beta R_0\} \exp\{2\beta |x_i|\}\leq  \exp\{-\beta R_0\}
\exp\{2\beta |X|\}$. Therefore,
\begin{align}
\Vert \widetilde{\Gamma}(N')\Vert ^2
&\geq \left(\Gamma(N'),
\ (1-\sum_i g_i)\Gamma(N')\right) \nonumber \\
&\geq 1-N'\exp\{-\beta R_0\}
\left(\Gamma(N'), \ \exp\{-2\beta|X|\} \Gamma(N')\right) \ . \nonumber
\end{align}

Together, (\ref{innerenergy}) and (\ref{normbound}) imply
(for $\exp\{\beta R_0 \} > N'\ C_\beta $)
\begin{align}\label{innerenergy2}
\frac{\left( \widetilde{\Gamma}(N'), H^V(N')\
 \widetilde{\Gamma}(N')\right) }
{\left(\widetilde{\Gamma}(N'),\ \widetilde{\Gamma}(N')\right)}
&\leq E^V(N') +
\frac{N'}{R_0^2}\  \frac{C_\beta}{ \exp\{\beta R_0 \} -N'\ C_\beta}
\nonumber \\
&\leq E^V(N') + \frac{2N'}{R_0^2}\  C_\beta e^{-\beta R_0}
 \  ,
\end{align}
where the last inequality holds provided that $R_0$ is chosen such that
$\beta R_0 \geq \log (2N' C_\beta)$.

The next step is to localize the photons in the state
$\widetilde{\Gamma}(N')$
in a ball centered at the origin of radius $L > R_0$. This leads to a new
state
$\Psi(N')$ with
\vfill \eject
\begin{multline}\label{localenergy2}
\frac{\left(\Psi(N'), H^V(N') \Psi(N')\right)} {\left(\Psi(N'),
\Psi(N') \right)} \\
\leq E^V(N') +  \frac{N'}{R_0^2}\  \frac{C_\beta}{
\exp\{\beta R_0 \} -N'\ C_\beta}
+\frac{CN'}{(L-2R_0)^{\gamma}} \frac{R_0}{L^{\gamma}} \ ,
\end{multline}
for all $\gamma <1$ and for $R_0$ and $L-2R_0$ large enough.

The construction of this function $\Psi$ is
precisely the same as the photon field localization that led to the
state $\Phi(n)$. It is carried out in Theorem \ref{locphot}.

Thus, $\Psi(N')$ is a state in which all the electrons are localized in a
ball
of radius $R_0$ and the photons are all localized in a ball of radius $L$.
Moreover the localization errors are small and given in
(\ref{localenergy2}).

\boxed{Part 3.} Now we put the pieces from Part 1 and Part 2 together and
construct a trial function $\Xi$ whose energy will be strictly below
$E^V(N')+E^0(n)$.

As mentioned above, since $H^0(n)$ is translation invariant we
can, by shifting, make sure that the photons in the state $\Psi(N')$
and the photons in the shifted state $\Phi(n)$,
live in disjoint sets. This will be the case when the smallest distance of
the
centers of the balls $B_1, \dots B_n$ with the center of the ball in which
$\Psi(N')$ lives is greater than  $2L$.

Now we can form the product state $\Xi$ as indicated in
(\ref{product}).  The state is symmetric in the photon variables by
construction. It has to be antisymmetrized in the electron labels
though, i.e., replace the products $\phi_{i_1,p_1; \cdots ;
i_m,p_m}\psi_{j_1,q_1; \cdots ; j_k,q_k}$ in (\ref{product}) by
\begin{multline}\label{product2}
c(N,N') \sum_{\pi \in S_N} (-1)^{\pi} \phi_{i_1,p_1; \cdots ;
i_m,p_m}(z_{\pi(1)},\cdots z_{\pi(N-N')}) \\ \times
\psi_{j_1,q_1; \cdots ; j_k,q_k}(z_{\pi(N-N'+1)}, \cdots , z_{\pi(N)}) \ .
\end{multline}
where $\pi$ runs through all the permutations of $N$ elements,
\begin{equation}\label{norm}
c(N,N')= \frac{1}{\sqrt{N!(N-N')!N'!}}
\end{equation}
is the normalization and $z_j=(x_j,s_j)$, the position and spin of the
$j$-th electron. The expression in (\ref{norm}) is
calculated by noting first that $\phi$ and $\psi$ are each
antisymmetric in their electron coordinates and, second, that there is
no overlap between $\phi$ and $\psi$ because the $x$ variables in the
two functions have disjoint support. Informally speaking, the
antisymmetrization in (\ref{product2}) has no effect and could be
dispensed with for all practical purposes since the operators $\nabla
$ and the potentials that we consider here are local operators.
If  we the Hamiltonian contained nonlocal operators, such as the
`relativistic' $\sqrt{-\Delta}$ then the antisymmetrization (\ref{product2})
would have a more profound effect --- although (\ref{norm}) is still
correct.

Next, we calculate (with $: \ \ : $ denoting normal ordering)
\begin{multline}
\left(\Xi, H(N) \ \Xi \right) \\
=  \left(\Xi, \sum_{i=1}^N \left[:(p_i + \sqrt{\alpha} A(x_i))^2:
+\frac{g}{2} \sqrt{\alpha}\> \sigma_i \cdot B(x_i)
+V(x_i)\right] \Xi\right) \\
+ \left(\Xi,\sum_{i <j}\frac{1}{|x_i-x_j|} \Xi\right)
+\left(\Xi, H_f \Xi\right)
\end{multline}
in terms of the normalized $ \Phi(n)$ and $\Psi(N')$. The field energy term
has been
explained previously
in equation (\ref{fieldproduct}) and yields
\begin{multline} \label{fieldlo}
\left(\Xi, H_f \Xi\right) = \left(\Phi(n), H_f  \Phi(n)\right)+
\left( \Psi(N'), H_f  \Psi(N')\right) \\
+ 8\pi\sum_i \Re \int \frac{\left(a_\lambda(x) \Phi(n),\Phi(n) \right)
\left( \Psi(N'), a_\lambda(y) \Psi(N') \right)}{ |x-y|^4}  dx dy  \\
+8\pi\sum_i \Re \int \frac{ \left( \Phi(n),a_\lambda(y)\Phi(n) \right)
\left(a_\lambda(x) \Psi(N'), \Psi(N')
\right)}{|x-y|^4} dx dy   \ .
\end{multline}
The Coulomb repulsion term is easily calculated to consist of three terms:
\begin{eqnarray} \label{coureplo1}
&&\left(\Phi(n), \sum_{N'<i <j\leq N} \frac{1}{|x_i-x_j|} \Phi(n)\right)
(\Psi(N'), \Psi(N')) + \\
&&\left(\Psi(N'), \sum_{1 \leq i <j\leq N'} \frac{1}{|x_i-x_j|}
\Psi(N')\right)
(\Phi(n), \Phi(n)) \label{coureplo2} + \\
&&\sum_{i=1,j=N'+1}^{N',N}  \int \frac{\Vert \Psi(N') \Vert^2
(x_1, .. , x_{N'} ) \Vert \Phi(n) \Vert^2(x_{N'+1}, .. , x_N)}
{|x_i-x_j|} d^Nx  \label{coureplo3} \ .\
\end{eqnarray}
Here the norm signs indicate that the norm has been taken in Fock space
and in the spin space.

The electron kinetic energy involves the calculation of terms of the form
\vfill\eject
\begin{multline}
\left(\Xi, a(f)^2 \Xi \right)\\  = \left(\Psi(N'), a(f)^2 \Psi(N')\right)
\left(\Phi(n), \Phi(n)\right) +
\left(\Phi(n), a(f)^2 \Phi(n)\right)\left(\Psi(N'), \Psi(N')\right) \\
+2 \Re  \left(\Psi(N'),a(f) \Psi(N')\right) \left(\Phi(n), a(f)
\Phi(n)\right) \ .
\end{multline}
Thus, we have that
\begin{align}
&\left(\Xi, \sum_{i=1}^N \left[:(p_i + \sqrt{\alpha} A(x_i))^2:
+\frac{g}{2}
\sqrt{\alpha}\> \sigma_i \cdot B(x_i)
\right] \Xi\right)  =    \nonumber \\
&\quad\sum_{i=1}^{N'}\left(\Psi(N'),  \left[:(p_i + \sqrt{\alpha} A(x_i))^2:
+\frac{g}{2}
\sqrt{\alpha}\> \sigma_i \cdot B(x_i)
\right] \Psi(N') \right) \left(\Phi(n), \Phi(n) \right)   \nonumber \\
&+\sum_{i=N'+1}^{N} \left(\Phi(n),  \left[:(p_i +
\sqrt{\alpha} A(x_i))^2: +\frac{g}{2}
\sqrt{\alpha}\> \sigma_i \cdot B(x_i)
\right] \Phi(n) \right) \left(\Psi(N'), \Psi(N') \right) \nonumber \\
&+\alpha \sum_{i=1}^{N'}\int
\left(\Phi(n), :A(x_i)^2: \Phi(n) \right)
\Vert\Psi(N')\Vert^2(x_1,\dots, x_{N'}) dx_1\cdots dx_{N'}
\label{kinlo1} \\
&+ \alpha\sum_{i=N'+1}^{N}\int \left(\Psi(N'), :A(x_i)^2: \Psi(N') \right)
\Vert\Phi(n) \Vert^2(x_{N'+1},.., x_N) dx_{N'+1} \cdot \cdot dx_N
\label{kinlo2} \\
&+2\alpha \sum_{i=1}^{N'}\int
\left(\Phi(n), A(x_i) \Phi(n) \right) \cdot
\left(\Psi(N'), p_i \Psi(N') \right)(x_1,\dots, x_{N'}) dx_1\cdots dx_{N'}
\label{pdota1} \\
&+ 2\alpha\sum_{i=N'+1}^{N}\int \left(\Psi(N'), A(x_i) \Psi(N') \right)
\cdot \nonumber \\
&\qquad\qquad\qquad\qquad \left(\Phi(n),
p_i \Phi(n) \right) (x_{N'+1}, \dots, x_N) dx_{N'+1}
\cdots dx_N \label{pdota2} \\
&+2\alpha \sum_{i=1}^{N}\int  \left\langle \Psi(N'), A(x_i) \Psi(N')
\right\rangle
\left\langle \Phi(n), A(x_i) \Phi(n) \right\rangle dx_1
\cdots dx_N  \label{kinlo3} \\
&+\frac{g}{2} \sqrt{\alpha}\>
\sum_{i=N'+1}^N \int \left(\Psi(N'), B(x_i)\Psi(N')\right) \cdot
\nonumber\\
&\qquad\qquad\qquad\qquad \left(\Phi(n),\sigma_i \Phi(n)\right) (x_{N'+1},
\dots,
x_N)dx_{N'+1}\cdots
dx_N
\label{kinlo4} \\
&+\frac{g}{2} \sqrt{\alpha}\>
\sum_{i=1}^{N'} \int \left(\Phi(n), B(x_i)\Phi(n)\right) \cdot \nonumber \\
&\qquad\qquad\qquad\qquad \left(\Psi(N'), \sigma_i
\Psi(N')\right) (x_1, \dots, x_{N'})
dx_1\cdots dx_{N'} \  .  \label{kinlo5}
\end{align}
Finally, the last and most important term
\begin{multline}
 \left(\Xi, \sum_{i=1}^N V(x_i) \Xi \right) =
\left(\Psi(N'), \sum_{i=1}^{N'} V(x_i)  \Psi(N')
\right)\left(\Phi(n), \Phi(n) \right)  \\
+\left(\Phi(n), \sum_{i=N'+1}^{N}  V(x_i)
\Phi(n) \right)\left(\Psi(N'), \Psi(N') \right)
\label{couattrlo} \ .
\end{multline}

Lemma \ref{asquare} allows us to show that the terms (\ref{kinlo1},
\ref{kinlo2}) are of order  $L^{-2\gamma}$ for any $\gamma <1$.
This follows from the
fact that in our trial function the electrons are localized in balls
of radius $R_0$, and so the distance $D$ in Lemma \ref{asquare}
between any electron and the localized photon field of the subsystem
to which the electron does {\it not } belong is at least
$L-R_0$. Since we can easily choose $L$ large and $R_0$ to be an
arbitrarily chosen small constant times $L$ we conclude, from Lemma
\ref{asquare}, that these terms are of order $L^{-2\gamma}$ for any $\gamma
<1$.

The terms (\ref{coureplo3},
\ref{couattrlo}) taken together are the terms that will give us
binding.  We shall show that, after averaging over rotations, the two
terms add to $-(Z-N')/3L$, which is less than $-{\rm pos.\ const.}/L$
according to our hypothesis.

{\it `Averaging over rotations'} means the following. We fix the state
$\Psi(N')$ of the electrons bound to the nuclei and their field, which
extends out a distance $L$ from the origin.  On the other hand, the
state of the $n$ unbound electrons was called $\Phi(n)$, but
actually there are infinitely many states we could use. That is, we
start with one $\Phi(n)$ and consider all rotations of it about the
origin. The average Coulomb interaction (i.e., the average of
(\ref{coureplo3}) and (\ref{couattrlo})) is the same as if the bound
electron state $\Psi(N')$, including the nuclei, was rotated about the
origin. However, the average potential generated by the latter average
over rotations at a point $x$ would be exactly $(Z-N')/|x|$ {\it
provided} $|x| > L$. This is Newton's theorem
\cite[Theorem 9.7]{Analysis}. Therefore, there exists a rotation
so that the Coulomb interactions (\ref{coureplo3}) and
(\ref{couattrlo}) are as if the inner state were a point charge
located at the origin and of strength $Z-N'$.

We now choose $\Phi(n)$ so that one of the balls of radius $L$ in
which the $n$ electrons and the field reside is tangent to the ball of
radius $L$ in which the bound electrons and field reside. By averaging
over rotations we may assume that the Coulomb potential seen by the $n$
electrons is that of a point charge at the origin; since there is at
least one of the outer balls that is a distance $2L$ from the origin,
and since that ball contains at least one electron, we can safely say
that the Coulomb interaction of the outer electrons with the nuclei,
i.e., the sum of the term (\ref{coureplo3}) and the last term in
(\ref{couattrlo}),
is less than $-(Z-N')/3L$. (The reason we wrote $Z-N'$ instead of
$n(Z-N')$ is that we do not know the positions of the other $n-1$
electrons; they could be very far away.)

To summarize the situation thus far, we have a negative Coulomb attraction
of order
$CL^{-1}$, where $C$ is a fixed constant.  We have various localization
errors of order
$L^{-2\gamma}$, $R_0^{-2}$ and also $(R_0/L^\gamma)
(L-2R_0)^{-\gamma}(1+|\log(\Lambda R_0)|)$.
These latter terms can be made arbitrarily small compared to $CL^{-1}$
if we choose $1> \gamma >3/4$  and
\begin{equation}
L^{2\gamma -1} >> R_0 >> L^{1/2} \ .
\end{equation}
 Finally, there
are the terms (\ref{fieldlo}) and (\ref{pdota1} -- \ref{kinlo5})  which
involve expectation values of linear operators $a^{\#}$ in $\Phi(n)$ and in
$\Psi(N')$.
These are dangerous looking terms; on the face of it they appear to possibly
be of order
$L^{-1}$, but we can make them all effectively vanish!

To eliminate these terms we can make an anti-unitary transformation on
$\Psi(N')$ (or
else on $\Phi(n)$, but not on both) that will not alter the energy of each
subunit or
alter the Coulomb interaction.  This anti-unitary is simply to replace
$a^{\#}$ by
$-a^{\#}$ and, simultaneously use complex conjugation to change $\Psi(N')$
to its complex
conjugate $\overline{\Psi(N')}$.  In addition we apply the unitary operator
$\mathcal{W}=\prod_{i=1}^{N'} \sigma_i^{(2)}$, where $\sigma^2$ is the
second Pauli matrix in the usual
basis in which $\sigma^2$ has purely imaginary elements and $\sigma^1$ and
$\sigma^3$ are
real.

The effect of applying this anti-unitary is to replace (\ref{fieldlo}) and
(\ref{pdota1} -- \ref{kinlo5}) by their negatives,
whereas all other energy terms
remain unchanged. Note that the anti-unitary when applied to $\Psi(N')$
changes
the sign of one of the factors in (\ref{pdota1} -- \ref{kinlo5}) only.
It changes the sign of $(\Psi(N'), p_i \Psi(N'))$
because of the complex conjugation and it changes the sign of
$(\Psi(N'), \sigma_i \Psi(N'))$  because of complex conjugation and
$\mathcal{W}$.
The terms $(\Psi(N'), A(x_i) \Psi(N'))$ and
$(\Psi(N'), B(x_i) \Psi(N'))$ change their sign because of the change of
sign of the $a^\#$'s.
Thus, each of this terms can be negated, and one choice or the other will
make
{\it the sum} (but perhaps not each individual term)  of (\ref{fieldlo}) and
(\ref{pdota1} -- \ref{kinlo5})
non-positive.
\hfill\qedsymbol


\section{Localization Estimates for  `Free'  Electrons and Photons}
\label{locestim}

The main result of this section is Theorem
\ref{locphot} which shows how to construct a state in which the `free'
electrons and the field are localized. This was used in Part 1 of the
proof of Theorem \ref{atom}. Part 2 of the proof of Theorem \ref{atom}
also uses the part of Theorem \ref{locphot} relating to the field
localization.

The proofs in this section rely, in part, on the commutator estimates
of Sects. \ref{commutators} and \ref{infraredbounds}.
The Hamiltonian for the $n$ free electrons is given in (\ref{free}).

\subsection{Localization of the Electrons, but not  the Photons}

\begin{lm}[localization of electrons]  \label{locelect}
Fix a radius $R_0>0$. Then there exist (not necessarily disjoint)
balls $B_1, \dots B_n$ in ${\mathbb R}^3$, each of radius $R_0$, and a
normalized vector $\Psi$ in the physical Hilbert space ${\mathcal H}(n)$
such
that the electronic part of $\Psi$ is supported in $\Omega(B_1, \dots
, B_n)$ and with an energy
\begin{equation}\label{electronlocenergy}
\left(\Psi, H^0(n) \Psi \right) < E^0(n) +  b\, n^2 R_0^{-2} ,
\end{equation}
where $b = 2\pi^2$ is twice the lowest eigenvalue of the Dirichlet
Laplacian in a ball of radius 1.
\end{lm}

{\it Conjecture:} The proof does not tell us the location of the $n$
balls. If they happen to be distinct then we can replace $n^2$ by $n$
in (\ref{electronlocenergy}). We conjecture that the theorem can be
generally improved in this way, i.e., $n^2\to n$, with, perhaps, a
different value for $b$.

\bigskip

{\it PROOF:} Let $\varepsilon = b\, n^2 R_0^{-2}/2$ and let $\Phi$ be a
normalized approximate ground state with error at most
$\varepsilon/2$, i.e., $\Phi \in {\mathcal H}$ and $\left(\Phi, H^0
\Phi\right) < E^0(n) +\varepsilon/2$. Let $B$ denote the ball of radius
$R_0$ centered at the origin in ${\mathbb R}^3$ and let $\chi$ be a
normalized, nonnegative, infinitely differentiable function with
support in $B$. Define the function $G$ of $X = (x_1, ..., x_n)$ and
$Y = (y_1, ..., y_n)$ by
\begin{equation}
G(X,Y) = \sum_{\pi \in S_n} \prod_{i=1}^n \chi(x_{i} - y_{\pi i}) \ ,
\end{equation}
where $S_n$ is the symmetric group.
Clearly $G$ is a symmetric function of the $X$ variables and of the
$Y$ variables and, therefore, $G(X,Y) \Phi$ is a valid vector in the
physical Hilbert space for each choice of $Y$.

It is obvious that \begin{equation}\label{perm}
P(X) := \int_{{\mathbb R}^{3n}} G(X,Y)^2 dY
\end{equation}
is simply $n !$ times the permanent of the $n\times n$ hermitian, positive
semidefinite matrix $M_{i,j} := \int_{{\mathbb R}^3}
\chi(x_i -y) \chi(x_j-y) dy$.  It is a general fact that such a
permanent is not less than the product of its diagonal elements; in
fact it is not less than the product of the permanent of any principal
$(n-m) \times (n-m)$ submatrix and the permanent of its $m\times m$
complement \cite{liebperm}, so $P(X) \geq n!$.  In particular, a fact
that we shall use later is that for each $i$, $P(X) \geq n! M_{i,i}
\widetilde{M}_{i,i}= n! \widetilde{M}_{i,i}$, where
$\widetilde{M}_{i,i}$ is the permanent cofactor of $M_{i,i}$, i.e., it
is the permanent of the matrix in which the $i^{\rm th}$ row and
column is deleted from $M$.  (In our case, the assertion is obvious
since every matrix element $M_{i,j} >0$.)

We define $W(X,Y) =  G(X,Y) P(X)^{-1/2}$, and using this
we define
\begin{equation}
\Psi_Y :=  W(X,Y) \Phi \ .
\end{equation}
Our $\Psi$ will be be $\Psi_Y$
(up to normalization) for a choice of $Y$  to be determined shortly.
Since $P(X) \geq n!$ the multiplier  $W(X,Y) $ is $C^\infty_c$.

We proceed analogously to Theorem 3.1 of \cite{GLL}. Consider
\begin{equation}\label{yineq}
\mathcal{E}(Y) := (\Psi_Y, H^0(n) \Psi_Y) -
\left[E^0(n) +\varepsilon + b \, n^2 R_0^{-2}\right]
\left( \Psi_Y, \, \Psi_Y\right)\ .
\end{equation}
Our goal is to show that $\int \mathcal{E}(Y) dY <0$ for a suitable
choice of $\chi$. This will prove that there is a set of $Y$'s of
positive measure such that $\Psi_Y \neq 0$ and also $(\Psi_Y, H^0(n)
\Psi_Y)/\left( \Psi_Y, \, \Psi_Y\right)\leq [E^0(n) +\varepsilon + b
\, n^2 R_0^{-2}]$, which is what we wish to prove.

It is obvious, from (\ref{perm}) that $ \int W(X,Y)^2  dY= 1$
and so
\begin{equation}
\int \left( \Psi_Y, \, \Psi_Y\right) dY = \left( \Phi, \, \Phi \right)=1.
\end{equation}
In a similar fashion one sees that
\begin{equation}
\int ( \Psi_Y, [\alpha \sum_{i<j} |x_i-x_j|^{-1} + H_f ]\Psi_Y ) dY
= ( \Phi, [\alpha \sum_{i<j} |x_i-x_j|^{-1} + H_f ] \Phi ) \ .
\end{equation}

Next, we compute
\begin{multline}
\Vert (\nabla_{x_i} +iA(x_i)) \Psi_Y \Vert^2 =
\Vert (\nabla_{x_i}W) \Phi \Vert^2
+ \Re\left(\Phi, (\nabla_{x_i} W^2)\cdot(\nabla_{x_i}+ iA(x_i) ) \Phi
\right) \\
+ \Vert W (\nabla_{x_i} + iA(x_i) ) \Phi \Vert^2 \ .
\end{multline}
The middle term vanishes when we integrate over $Y$ since $ \int
W(X,Y)^2 dY= 1$ and hence  $ \int \nabla_{x_i} W(X,Y)^2 dY= 0$. The
last term gives us the required contribution of the kinetic energy to
$\int (\Psi_Y, H^0(n) \Psi_Y)dY$ in (\ref{yineq}), again using the
fact that $ \int W(X,Y)^2 dY= 1$\ .  The first term is $\left( \Phi,
F_i(X) \Phi \right)$, where
\begin{equation}
F_i(X) = \int \left|\nabla_{x_i}\left\{ G(X,Y)P(X)^{-1/2}\right\}
\right|^2 dY \ .
\end{equation}

Our proof is complete if we can show that
$F_i(X) \leq 3\varepsilon/2n$,
which we shall do next. We start with
\begin{multline}
\nabla_{x_i}\left\{G(X,Y)P(X)^{-1/2}\right\} = P(X)^{-1/2}
\nabla_{x_i} G(X,Y)\\ -(1/2) G(X,Y) P(X)^{-3/2}
\nabla_{x_i} P(X) \ .
\end{multline}
If we square this and integrate over $Y$ we obtain
(recalling that $\nabla_{x_i} P(X) =2\int G(X,Y) \nabla_{x_i}G(X,Y) dY$ and
$\int G(X,Y)^2 dY =P(X)$ )
\begin{equation}
F_i(X) = \frac{1}{P(X)}\int |\nabla_{x_i} G(X,Y)|^2 dY
-\frac{1}{4 P(X)^2}|\nabla_{x_i} P(X)|^2  \ .
\end{equation} We shall ignore the last term since it is
negative.

In order to  compute $\nabla_{x_i} G(X,Y)$ let us write
\begin{equation}
G(X,Y) = \sum_{j=1}^n \chi (x_i -y_j) \mu_j (X', Y') := \sum_{j=1}^n
a_j(X,Y),
\end{equation}
where
\begin{equation}
 \mu_j (X', Y') = \sum_{\pi \in S_{n-1}}
\prod_{\ell \neq i}
\chi\left(x_\ell - y_{\pi \ell}\right) \ ,
\end{equation}
and where $S_{n-1}$ denotes  the set of bijections of
$1, .\ .\ ,\hat{i}, .\ .\ , n$
into $1, .\ .\ ,\hat{j}, .\ .\ , n$.
Then,
\begin{multline}
\int \left|\nabla_{x_i} G \right|^2 dY = \sum_{j=1}^n \sum_{k=1}^n
\int \nabla_{x_i}a_j\cdot
\nabla_{x_i}a_k dY
\leq  n\sum_{j=1}^n\int \left|\nabla_{x_i} a_j \right|^2 dY\\
=n\sum_{j=1}^n \int\left|(\nabla \chi)(x_i-y_j)\right|^2 dy_j
\int \mu_j(X', Y')^2 dY'
=n  C_{i,i} \int |\nabla \chi(x) |^2 dx \ ,
\end{multline}
where $Y'=(y_1, \dots, \hat{y_j}, \dots, y_n)$ and where $C_{i,i} =
\int \mu(X', Y')^2 dY'$ equals $(n-1)!$ times $\widetilde{M}_{i,i}$,
the cofactor of $M_{i,i} $ in the permanent of $M$.  However, $P(X)
\geq n! M_{i,i}\widetilde{M}_{i,i} = n!
M_{i,i}\left[C_{i,i}/(n-1)!\right] $, as explained before. Therefore,
$P^{-1}\int \left|\nabla_{x_i}G\right|^2 dY \leq n \int |\nabla
\chi|^2$.  The same inequality holds for any $i=1, \dots , n$ which
gives us a factor of $n^2$ altogether.

At this point we wish to choose $\chi$ to be the lowest Dirichlet
eigenfunction of $-\Delta$ in the ball $B$. That function is not in
$C^\infty_c(B)$, but it can be approximated by such a function so that
the error  in $n^2 \int |\nabla \chi|^2$
is less than $\varepsilon / 2$.
\hfill \qedsymbol

\subsection{Localization of Photons}

Our next task is to produce a state in which the photons are
localized.  A definition is needed first. We say that the
electromagnetic field in a state $\Phi$ is {\bf supported} in a closed
subset $\Sigma \subset \R^3$ if each component $a_\lambda$ of the
field satisfies $\Vert a_\lambda(x) \Phi\Vert =0$ for all $x \notin
\Sigma$.  To construct a localized state from any given state $\Phi$
we use the representation (\ref{rep}) of Fock space.  We suppress the
space and spin variables of the electrons for the moment. For a smooth
cutoff function $0\leq j(y)\leq 1$ define the localization operator
$\J$ on Fock space in the following manner. $\J \Phi$ is still given
by (\ref{rep}) but the vector $|i_1, m_1;\ \cdots \ ;i_n,m_n\rangle$
is changed to
\begin{equation}\label{jay}
\J|i_1, m_1;\ \cdots \ ;i_n,m_n \rangle
=\frac{1}{\sqrt{m_1! \cdots m_n!}}
\ a^*(jf_{i_1})^{m_{1}} \cdots a^*(jf_{i_n})^{m_{n}} |0\rangle \ ,
\end{equation}
Clearly, $\J$ is a linear, self-adjoint operator and
\begin{equation}\label{localjay}
\Vert a_\lambda(y) \J\Phi \Vert = 0
\end{equation}
for all $y$ that are outside the support of the function $j(y)$.
Note that $\J$ is a contraction, i.e., $\Vert \J \Phi\Vert \leq
\Vert \Phi \Vert $ for
all $\Phi$.

To work effectively with the operator $\mathcal{J}$
the following commutation relations will be useful later.  Since
the electron variables are not
relevant for the calculation, we suppress them here.

\begin{lm}[Commutation relations for $\mathcal{J}$]\label{jaycomrel}
For any $f$ in the single photon Hilbert space $L^2(\R^3; \C^2)$ we have
(with $a_\lambda (f) =\int a_\lambda(y) \overline{f(y)} dy $, as before)
that
\begin{align}
a_\lambda (f)\J&= \J a_\lambda(jf) ,&\J a_\lambda^*(f) &= a_\lambda^*(jf) \J
\\
\left[a_\lambda(f),\J \right] &= \J a_\lambda((j-1)f) ,&
\left[a_\lambda^*(f),
\J\right] &=-a_\lambda^*((j-1)f)\J \ .
\end{align}
\end{lm}
\medskip
{\it PROOF:}
For any state $\Psi$ in the Fock space we have that
\begin{multline}
\left[a_\lambda(f) \J \Psi\right]_n(y_1,\lambda_1; \dots ; ,y_n \lambda_n)\\
=\sqrt {n+1} \prod_{k=1}^{n} j(y_k)
 \int \overline{f(y)}j(y)
\left[\Psi \right]_{n+1}(y ,\lambda; y_1,\lambda_1; \dots; y_{n},\lambda_n)
d y \\
= \left[\J a_\lambda(jf) \Psi\right]_n(y_1,\lambda_1; \dots ;y_n, \lambda_n)
\ .
\end{multline}
All the relations follow immediately from this.
\hfill\qedsymbol

\subsection{Localization of the Photons and the Electrons Together}

One would like to think that most of the photons ought to be localized
near the electrons. This will be the case provided one replaces the
state $\Psi$ of Lemma \ref{locelect} by the ground state for the
Hamiltonian $H^0(n)$ restricted to the states that vanish outside the
set $\Omega$. Moreover, this state will have an energy close to the
energy $E^0(n)$.  The following theorem makes this precise.

\begin{thm}[Localized photons and free electrons]  \label{locphot}
Fix radii $R_0>0$ and $L >2R_0$. Then there exist (not necessarily
disjoint) balls $B_1, \dots B_n$ in ${\mathbb R}^3$, each of radius
$R_0$ and a normalized vector $\Phi (n)$ in the physical Hilbert space
${\mathcal H}(n)$ such that the electronic part of $\Phi (n)$ is supported
in $\Omega(B_1, \dots , B_n)$ and the electromagnetic field is
supported in $\Sigma = \cup_{i=1}^n P_i$ where $P_i$ is a ball
concentric with $B_i$ but with radius $L$.

The energy of $\Phi (n) $ satisfies
\begin{equation}\label{locenergy}
\left(\Phi (n), H^0(n) \Phi (n) \right) < E^0(n) +  b\, \frac{n^2}{ R_0^{2}}
+ c \frac{n}{ (L-2R_0)^{\gamma}}
\left(\frac{R_0}{L^\gamma}\right)(1+|\log(\Lambda R_0)|)    ,
\end{equation}
for any $\gamma < 1$ and where $b = 2\pi^2$ is twice the lowest
eigenvalue of the Dirichlet Laplacian in a ball of radius 1. The
constant $c$ depends only on $\gamma$ and  is independent of $R_0$
and $L$.

\end{thm}

{\it PROOF:} We start with the wave function $\Psi$ given by Lemma
\ref{locelect}. This fixes the balls $B_1, \dots B_n$, and hence the
symmetrized product $\Omega(B_1, \dots B_n)$.  The next step is to
redefine the Hamiltonian $H^0(n)$ by restricting the Hilbert space to
the balls, i.e., we replace the space $\wedge L^2(\R^3; \C^2) $ by the
subspace
of $L^2(\Omega; \otimes_1^n \C^2)$ consisting of functions that are
antisymmetric under the exchange of particle labels. (This makes sense
because $\Omega$ is symmetric under exchange of particle coordinates.)
The Laplacian is replaced by the Dirichlet Laplacian.

A physical way to say this is that we add an infinite potential
outside $\Omega$. This is not a sum of single particle potentials, but
that is immaterial. The point is that by the methods of \cite{GLL}
there is a bound state, i.e., there is a state $\Phi_D(n)$ with lowest
energy $E^0_D(n)$ (the letter D stands for `Dirichlet') that satisfies
Schr\"odinger's equation. (In fact, the methods of \cite{GLL} are not
needed to establish the existence of a ground state in this case since
all finite energy states are evidently localized; this was noted
earlier in \cite{gerard},
\cite{BFS} and \cite{hirokawa}. However, \cite{GLL} is needed for the
photon localization  in the next step.)

This ground state will obviously have a lower energy than the $\Psi$
given by Lemma \ref{locelect} since that $\Psi$ automatically
satisfies the Dirichlet boundary conditions, that is
$\left(\Phi_D(n), H^0(n)\Phi_D(n)\right) \leq (E^0(n) +  b\, n^2 R_0^{-2} )
\left(\Phi_D(n), \Phi_D(n)\right)$.

Next, we localize the photons in the set $\Sigma$. A standard IMS
localization yields two smooth functions $j_1(y), j_2(y)$ with
\begin{equation}\label{j1j2}
j_1(y)^2+j_2(y)^2=1
\end{equation}
with support of $j_1(y)$ in $\Sigma$, We also require that $j_1(y)$ is
identically equal to $1$ on the set $\cup_{i=1}^n Q_i$ where, for each
$i$, $Q_i$ is a ball of radius $L/2$, concentric with $P_i$. Moreover
we can assume that $|\nabla j_i(y)| \leq C/L$ for some constant $C$
and $i=1,2$.  We define $\J\Phi_D(n)$ by using $j_1$ in (\ref{jay})
and, with the help of (\ref{localjay}), we use the localized state
$\J\Phi_D(n)$, appropriately normalized, as a trial function.
This function will be the required function $\Phi(n)$ of our
theorem.

The energy of the state $\J \Phi_D(n)$ can be compared
with the energy of $\Phi_D(n)$ by using the commutator formula
\begin{equation}\label{commutatorenergy}
\left(\J\Phi_D(n), (H^0(n)- E^0_D(n)) \J \Phi_D(n) \right)
=\left(\J\Phi_D(n),[H^0(n),\J]\Phi_D(n)\right) \ .
\end{equation}

 An important point about having a {\it ground state} $\Phi_D(n)$ is that
one can derive infrared bounds for this state (see
Sects. \ref{commutators} and \ref{infraredbounds}).  All that is needed is
that  $\Phi_D$
is a ground state, i.e., it satisfies the Schr\"odinger equation in order to
apply
the `pull through formula'.

Lemma \ref{normalization} shows that
the norm $\Vert \J\Phi_D \Vert$ is close to one and Lemma
\ref{loccomm} shows that the right side of (\ref{commutatorenergy}) is
bounded as
\begin{multline} \label{combound}
\left(\J\Phi_D(n),[H^0(n),\J]\Phi_D(n)\right) \\ \leq
\frac{C}{(L-2R_0)^{\gamma}}\left(\frac{R_0}{L^\gamma}\right)(1+|\log(\Lambda
R_0)|)
\left(\J\Phi_D(n),\J\Phi_D(n)\right)
\ ,
\end{multline}
for any $\gamma < 1$, where the constant $C$ depends on $\gamma$ but not on
$R_0$ and $L$.
This shows that $\Phi (n)= \J\Phi_D(n) / \Vert \J\Phi_D(n) \Vert$ satisfies
(\ref{locenergy}).
\quad\quad \quad  \hfill \qedsymbol

\section{Commutator and Related Estimates} \label{commutators}

In this section we prove various results stated in the previous
sections, particularly Lemma \ref{loccomm} which is used in the proof
of Theorem \ref{locphot}. We also prove Lemma \ref{asquare},
which is not a commutator estimate; it is simpler.
It is needed to bound  the terms (\ref{kinlo1}) and (\ref{kinlo2}),
as we stated just after eq. (\ref{couattrlo}).

We will deal mainly with the Dirichlet
ground state associated with electrons localized in the set
$\Omega(B_1, \dots B_n)$. The bounds for the ground state
describing electrons exponentially localized near the nuclei is easier
and follows in the same fashion.

In this section and the next we denote the Dirichlet ground state
$\Phi_D(n)$ simply by $\Phi$ in order to simplify the notation.

Recall the definitions of $j_1, j_2$ in (\ref{j1j2}) and of the
operator $\mathcal{J}$ in (\ref{jay}) which is defined by substituting
$j_1$ for $j$. An important operator in our analysis is the outer
photon number, given by
\begin{equation}
 \mathcal{N}_{\rm out} = \sum_{\lambda=1}^2
\int_{j_2(y)>0}  a_{\lambda}^*(y)a_{\lambda}^{\phantom{*}}(y)  d y
\end{equation}
or, in terms of matrix elements,
\begin{equation}
\left(\Phi, \mathcal{N}_{\rm out} \Phi\right):= \sum_{\lambda=1}^2
\int_{j_2(y)>0} \Vert a_{\lambda}(y) \Phi \Vert^2 d y
\end{equation}

We start with the following bound, which is
a consequence of the infrared bounds  proved in Sect. \ref{infraredbounds}.
It is used in the proofs of Lemmas \ref{normalization}, \ref{kincomlem} and
\ref{fieldcom}.

\begin{lm}[Photon number is small far away from the electrons]
\label{outerphotons}
For the Dirichlet ground state $\Phi$ of the free electrons localized in
$\Omega(B_1, \dots B_n)$ we have the bound (with $C$ independent of $R_0$
and
$L$)
\begin{equation}
\left(\Phi, \mathcal{N} \Phi\right) \leq C(1+|\log(\Lambda R_0)|) \ ,
\label{photonnumber}
\end{equation}
 and for all $\gamma < 1$, the bound
\begin{equation}
\left(\Phi, \mathcal{N}_{\rm out} \Phi\right)  \leq
C\left(\frac{R_0}{L^\gamma}\right)^{2} \Vert \Phi \Vert ^2
\end{equation}
where the constant $C$ depends on $n, \Lambda, \gamma$  but not on $R_0$ and
$L$.
Likewise, for the ground state $\Psi$ of the bound system given by the
Hamiltonian $H^V(N')$
we have that (with $C$ independent of $R_0$ and $L$)
\begin{equation}
\left(\Psi, \mathcal{N} \Psi \right) \leq C \ ,
\end{equation}
and for all $\gamma <1$ and $L>2R_0$ that
\begin{equation}
\left(\Psi, \mathcal{N}_{\rm out} \Psi \right) \leq
C\left(\frac{1}{L^{2\gamma}}\right) \Vert \Psi \Vert ^2  \ ,
\end{equation}
where the constant $C$ depends only  on $\gamma$.
\end{lm}

Our goal is to prove inequality (\ref{combound}), which is Lemma
\ref{loccomm} of this section. To do so, we shall need the following
three lemmas, in which $\J$ is defined with $j_1$. Recall that $0\leq j_1(y)
\leq 1$ and $j_1(y) =1$ for $y\in \bigcup_{i=1}^n Q_i$ (see the proof of
Theorem \ref{locphot}).

\begin{lm} \label{normalization}
For the normalized ground state $\Phi$, we have for all $0 < \gamma<
1$ that
\begin{equation}
1-\Vert \J \Phi \Vert^2 \leq C \left(\frac{R_0}{L^\gamma}\right)^{2}
\end{equation}
where $C$ is a constant that depends on $\gamma, n, \Lambda$ but not on
$R_0$ and $L$. Moreover, for an arbitrary state $\Psi$,
\begin{equation} \label{anystate}
\left(\J \Psi, \mathcal{N}_{\rm out} \ \J \Psi \right) \leq
\left( \Psi, \mathcal{N}_{\rm out} \ \Psi \right) \ .
\end{equation}
\end{lm}

{\it PROOF:} Formula (\ref{anystate}) is immediate from
\begin{equation}
\left(\mathcal{J} \Psi, \mathcal{N}_{\rm out} \mathcal{J} \Psi \right)
=\sum_{n=1}^\infty n \Vert \prod_{l=1}^n j(y_l)  \prod_{l=1}^n
\chi_{j_2(y_l)>0}[\Psi]_n \Vert^2  \ . \label{sequences}
\end{equation}

Next, note that
\begin{equation}\label{induct}
1-\prod_{k=1}^n j_1^2(y_k) = \sum_{l=1}^{n} \prod_{k=1}^{l-1}
j_1^2(y_k) j_2^{\> 2}(y_l)
\end{equation}
(by definition the empty product equals $1$). This is proved by
inserting $j_2^{\> 2}(y_n) = 1- j_1^2(y_n) $ on the left side of
(\ref{induct}) and then repeating the process inductively.  In
particular, we have that
\begin{equation}
1-\prod_{k=1}^n j_1^2(y_k) \leq \sum_{l=1}^{n} j_2^{\> 2}(y_l) \ ,
\end{equation}
from which we obtain
\begin{equation}
1-\Vert \J \Phi \Vert^2 \leq \left( \Phi, \mathcal{N}_{\rm out} \Phi \right)
\leq C \left(\frac{R_0}{L^\gamma}\right)^{2}
\end{equation}
by Lemma \ref{outerphotons}.
\hfill\qedsymbol

\begin{lm}  \label{kincomlem}
For the ground state $\Phi$ and for
every $L > 2 R_0$ we have that
\begin{multline}\label{kincom}
\left|\left(\J \Phi, \sum_{i=1}^n\left[(p_i+A(x_i))^2, \J \right]
\Phi \right)\right|\\ \leq \
C \frac{1}{ (L-2R_0)^{\gamma}}    \left(\frac{R_0}{L^\gamma}\right)
(1+|\log(\Lambda R_0)|)
\Vert \J \Phi \Vert^2\ ,
\end{multline}
for all $\gamma <1$. $C$ is a constant that depends on $\gamma, n, \Lambda$
but not on $R_0$ and $L$.
(Note that it makes no difference whether we use
$(p_i+A(x_i))^2$ or use its normal ordering since the
commutator of $a_i$ and $a_i^*$ is proportional to the identity
operator, which commutes with $\J$.)
\end{lm}
\medskip
{\it PROOF: }  We first calculate the commutator of $(p+A(x))^2$ with
$\J$.
\begin{multline}\label{terms}
\sum_{i=1}^n \left[(p_i+A(x_i))^2, \J \right]= \\
\sum_{i=1}^n  \left( 2p_i\cdot([a_i,\J]+[a_i^*, \J])
+ [a_ia_i,\J] + [a_i^* a_i^*, \J] + 2 [a_i^*a_i, \J] \right) \ ,
\end{multline}
where we abbreviated $a_\lambda (h^j_\lambda (x_i-\cdot))$ by $a_i$
and likewise for $a_i^*$.(Note that the index $j \in \{1,2,3\}$ of the
coupling function is unimportant and will be suppressed from now on.)

{\it Step 1.} The term $\sum_{i=1}^n \left( \J \Phi, 2p_i \cdot [a_i,\J]
\Phi \right)$
is bounded, by Schwarz's inequality, by
\begin{equation}\label{fivefifteen}
2\left(\sum_{i=1}^n \Vert p_i\J \Phi \Vert ^2 \right)^{1/2}
\left(\sum_{i=1}^n \Vert [a_i,\J] \Phi \Vert^2\right)^{1/2}\ .
\end{equation}
The first factor can be estimated simply in terms of the energy while
the second factor will deliver the necessary decay in $L$.  Using
Lemma \ref{jaycomrel} the problem is reduced to estimating, for each
fixed $X=(x_1, \dots, x_n) \in \Omega$ and each $i$ (with $\Vert \cdot
\Vert$ denoting the norm in Fock space only)
\begin{multline}\label{comcom}
\Vert \J a(h(x_i-\cdot)(j_1(\cdot)-1)) \Phi \Vert \\ \leq
\Vert a(h(x_i-\cdot)(j_1(\cdot)-1)) \Phi \Vert
\leq \int (1-j_1(y))\  \Vert h(x_i - y) a(y) \Phi \Vert d y \\
\leq \int (1-j_1(y))\ |h(x_i - y)|
\Vert  a(y) \Phi \Vert d y\\
\leq \left(\int (1-j_1(y))\ |h(x_i - y)|^2 d y\right)^{1/2}
\left( \int  (1-j_1(y))\Vert  a(y) \Phi \Vert^2 d y \right)^{1/2} \ .
\end{multline}
The first factor in (\ref{comcom}) can
be bounded, using Lemma \ref{coupling}, by
\begin{equation}
\left(\int  (1- j_1(y))\ \frac{1}{(L-2R_0)^{2\gamma}}
|x_i-y|^{2 \gamma}|h(x_i - y)|^2  d y \right)^{1/2}
\leq \frac{C }{( L-2R_0)^{\gamma}}
\end{equation}
since, whenever $ 1-j_1(y) \neq 0$, the distance between $x_i$ and $y$
is at least $d=(L/2)-R_0$, by construction. If we now square
(\ref{comcom}) and integrate over
$X$ we get the desired decay estimate for (\ref{fivefifteen}).

For the second factor in (\ref{comcom}) we note
that $1-j_1(y) \leq j_2(y)^2$. This, together with Lemma
\ref{outerphotons}, yields
\begin{equation}\label{final}
\left( \sum_{i=1}^n \Vert \J a(h(x_i-\cdot)(j_1(\cdot)-1)) \Phi
\Vert^2 \right)^{1/2}\leq \frac{C}{ (L-2R_0)^{\gamma}}
\left(\frac{R_0}{L^\gamma}\right)
\end{equation}
for all $\gamma <1$.

Next, $\left( \J \Phi, 2\sum_{i=1}^n p_i \cdot [a_i^*,\J] \Phi
\right)= -2\sum_{i=1}^n \left( [a_i,\J]\J \Phi, p_i \Phi
\right)$ and  this can be estimated in the same fashion as before except
that the estimate is in terms of
$$
\left( \J \Phi, \mathcal{N}_{{\rm out}} \J \Phi \right)
$$
instead of $\left( \Phi, \mathcal{N}_{{\rm out}} \Phi \right)$.
On account of Lemma \ref{normalization}, this is bounded by
$$
\left( \Phi, \mathcal{N}_{{\rm out}}  \Phi \right) \ .
$$
Hence, we obtain the same kind of bound as in (\ref{final}), i.e.,
for all $\gamma <1$,
$$
|\sum_{i=1}^n \left( \J \Phi, 2p_i\cdot[a_i^*, \J] \Phi\right)|
\leq\frac {C }{ (L-2R_0)^{\gamma}}  \left(\frac{R_0}{L^\gamma}\right) \ .
$$

{\it Step 2.}
Returning to (\ref{terms}) we concentrate on the term $[a_ia_i,\J]$
 which can be written as
$a_i[a_i,\J]+[a_i,\J]a_i$. Using Schwarz's inequality
\begin{equation}
\left(\J \Phi, a_i[a_i,\J] \Phi \right) = \left(a_i^* \J \Phi, [a_i,\J]
\Phi \right) \leq
\Vert a_i^* \J \Phi \Vert \ \Vert [a_i,\J] \Phi \Vert  \ .
\end{equation}
The
second factor is treated in precisely the same fashion as in
Step 1. The first factor cannot be estimated directly in terms of
the energy, since the function $\mathcal{J} \Phi$ is not an eigenfunction.
This will be dealt with below where we estimate the term
$\Vert a^* \mathcal{J}^2 \Phi \Vert$.

The term
\begin{equation}
\left( \J \Phi, [a_i,\J]a_i \Phi \right)  \ ,
\end{equation}
can be written, using Lemma \ref{jaycomrel}, as
\begin{multline}
\left( \J \Phi, \J a(h(x_i-\cdot)(j_1(\cdot)-1)) a(h(x_i-\cdot)) \Phi
\right)\\
= \left( \J^2 \Phi, a(h(x_i-\cdot)) a(h(x_i-\cdot)(j_1(\cdot)-1))  \Phi
\right)
\end{multline}
which, again using Schwarz's inequality, can be bounded by
\begin{equation}
\Vert a^*(h(x_i-\cdot)) \J^2 \Phi \Vert \ \Vert
a(h(x_i-\cdot)(j_1(\cdot)-1))
\Phi \Vert \ .
\end{equation}
As before, the first factor cannot be estimated in terms of the energy,
since $\mathcal{J}^2\Phi$
is not an eigenstate.
Note, however,  that
\begin{equation}
\Vert a^* \mathcal{J}^2 \Phi \Vert ^2 = \Vert a \mathcal{J}^2 \Phi
\Vert ^2 +\Vert h \Vert^2
\Vert \mathcal{J}^2 \Phi \Vert ^2 \ .
\end{equation} and the first term on the right side can be estimated by
\begin{equation}
\Vert h \Vert^2 \left(\mathcal{J}^2 \Phi , \mathcal{N} \mathcal{J}^2
\Phi\right)
\leq  \Vert h \Vert^2 \left(\Phi , \mathcal{N} \Phi\right)  \ .
\end{equation}
This follows from the formula
\begin{equation}
\mathcal{N} = \sum_{j=1}^\infty a^*(f_j)a(f_j) \ ,
\end{equation}
which is valid for any orthonormal basis $\{f_j\}$, in which we pick
$f_1(y) = h(x-y)/ \Vert h(x-\cdot) \Vert$, and from (\ref{sequences}).
(Here $h$ is an abbreviation for the coupling functions.)
Thus, using Lemma \ref{outerphotons},
\begin{equation}
|\sum_{i=1}^n \left( \J \Phi, [a_ia_i,\J] \Phi \right) | \leq C
\frac{1 }{ (L-2R_0)^{\gamma}}  \left(\frac{R_0}{L^\gamma}\right)
(1+|\log(\Lambda R_0)|)      \ .
\end{equation}

{\it Step 3.}
By taking adjoints the  third term in (2), $\sum_{i=1}^n [a_i^*a_i^* ,\J]$
leads
to the expression
\begin{equation}
-\sum_{i=1}^n\left( [a_ia_i ,\J] \J \Phi, \Phi \right)
\end{equation}
and can be dealt with in the same fashion as in Step 2.
It remains to analyze  $\sum_{i=1}^n [a_i^*a_i, \J]=
\sum_{i=1}^n a_i^*[a_i, \J] + [a_i^*, \J] a_i$.
The first term is estimated using
\begin{equation}
\left( a_i \J \Phi, [a_i , \J] \Phi \right) \leq \Vert  a_i\J \Phi \Vert \
\Vert [a_i, \J] \Phi \Vert \ ,
\end{equation}
while the second one can be written as
\begin{equation}
-\left( [a_i, \J] \J \Phi, a_i \Phi \right)
\end{equation}
which, once more  by Schwarz's inequality, can be bounded by
\begin{equation}
\Vert [a_i, \J] \J \Phi \Vert \ \Vert a_i \Phi  \Vert \ .
\end{equation}
Both these terms have been estimated previously.
\hfill\qedsymbol

We come now to the third lemma needed for the proof of Lemma
\ref{loccomm}.  This lemma concerns only the real part of a commutator
expectation value (\ref{commutatorexpression}), but this is all we
need for Lemma \ref{loccomm}.  The reason is that the total commutator
in Lemma \ref{loccomm} is manifestly real, since $\Phi(n)$ is an
eigenstate of $H^0(n)$ and $\J$ is selfadjoint.  On the other hand,
the piece of the commutator considered in Lemma \ref{kincomlem} is also
manifestly real and the only other part of $H^0(n)$ to be considered
is the potential energy terms, which commute with $\J$. Therefore, the
commutator expectation value in (\ref{commutatorexpression}) is, in
fact, real. The proof of Lemma \ref{fieldcom} is greatly simplified,
however, by being able to ignore the (non-existent) imaginary part.

\begin{lm}[Commutator of $\J$ with the field energy] \label{fieldcom}
The ground state $\Phi$ satisfies the bound
\begin{equation}\label{commutatorexpression}
\Re \left( \J \Phi, \left[H_f, \J\right] \Phi \right) \leq
\frac {C} { L}    \left(\frac{R_0}{L^\gamma}\right)(1+|\log(\Lambda
R_0)|)^{1/2}
\end{equation}
for any $\gamma < 1$ . $C$ is a constant that depends on $\gamma, n,
\Lambda$ but not on $R_0$ and $L$.
\end{lm}
\medskip

{\it PROOF:} It is convenient to write the field energy of a state $\Phi$ in
the form
\begin{multline}
\left(\Phi, H_f \Phi \right)=(2\pi)^3 \sum_\lambda \left(a_\lambda
(\cdot)\Phi, \sqrt{-\Delta}\
a\lambda (\cdot) \Phi \right) =\\
4\pi \sum_\lambda\int \frac{\Vert a_\lambda (x)\Phi - a_\lambda (y) \Phi
\Vert^2 }
{ |x-y|^4 } d x d y \ ,
\label{integralexpression}
\end{multline}
(see \cite[Eq. 7.12(4)]{Analysis} ).  Next,
we note that the commutator expression (\ref{commutatorexpression}) is
given by
\begin{multline} \label{remainderjunk}
\left( \J \Phi,H_f \J \Phi \right)-\Re \left( \J \Phi, \J H_f \Phi \right)
= 4\pi \sum_\lambda\int \frac{\Vert a_\lambda (x)\J \Phi - a_\lambda (y)
\J \Phi \Vert^2 }
{ |x-y|^4 } d x d y \\
-4\pi \sum_\lambda \Re\int \frac{\left(a_\lambda (x)\J^2 \Phi - a_\lambda
(y) \J^2
\Phi,a_\lambda (x) \Phi - a_\lambda (y) \Phi \right)
  }{ |x-y|^4 } d x d y \ .
\end{multline}

First, we investigate the numerator of the sum of the two integrands,
which is
\begin{multline}\label{numerator}
\left(a_\lambda (x)\J \Phi,a_\lambda (x)\J \Phi \right)-
\Re\left(a_\lambda (x)\J^2 \Phi,a_\lambda (x)\Phi \right)\\
-\Re\left(a_\lambda (x)\J \Phi,a_\lambda (y)\J \Phi \right)
+\Re\left(a_\lambda (x)\J^2
\Phi,a_\lambda (y)\Phi \right)
\end{multline}
plus the same thing with $y$ and $x$ exchanged. For brevity's sake we
have omitted the sums over $\lambda$ here and in the following. We
note that the $x$-$x$ terms (and likewise the $y$-$y$ terms) cancel.
This follows from
\begin{multline}
\left(a_\lambda (x)\J \Phi,a_\lambda (x)\J \Phi \right) -\Re
\left(a_\lambda (x)\J^2
\Phi,a_\lambda (x)\Phi \right) = \\
\left(a_\lambda (x)\J \Phi,a_\lambda (x)\J \Phi \right) -\Re
\left(\J a_\lambda (x)\J \Phi,a_\lambda (x)\Phi \right)
- \Re \left(\left[a_\lambda (x),\J\right] \J \Phi,a_\lambda (x)\Phi \right)
\\
=\Re \left(a_\lambda (x)\J \Phi, \left[a_\lambda (x),\J\right] \Phi \right)-
\Re \left(\left[a_\lambda (x),\J\right] \J \Phi,a_\lambda (x)\Phi \right) \
,
\end{multline}
which, together with Lemma \ref{jaycomrel}, yields
\begin{multline}
(j_1(x)-1)\Re \left(a_\lambda (x)\J \Phi, \J a_\lambda (x) \Phi \right)-
(j_1(x)-1)
\Re \left(\J a_\lambda (x) \J \Phi,a_\lambda (x)\Phi \right) \\
=(j_1(x)-1)\left[\Re \left(a_\lambda (x)\J \Phi, \J a_\lambda (x) \Phi
\right)-
\Re \left( a_\lambda (x) \J \Phi, \J a_\lambda (x)\Phi \right)\right] = 0 \
.
\end{multline}

Now we deal with last two terms in (\ref{numerator}).
\begin{multline}
\Re\left(a_\lambda (x)\J^2 \Phi,a_\lambda (y)\Phi \right)
-\Re\left(a_\lambda (x)\J \Phi,a_\lambda (y) \J \Phi \right) \\
=\Re \left(\left[a_\lambda (x),\J \right] \J \Phi,a_\lambda (y)\Phi \right)
+\Re
\left(a_\lambda (x)\J \Phi, \J a_\lambda (y)\Phi \right)\\
-\Re \left(a_\lambda (x)\J \Phi,a_\lambda (y)\J \Phi \right) \\
=\Re \left(\left[a_\lambda (x),\J \right] \J \Phi,a_\lambda (y)\Phi \right)
+\Re
\left(a_\lambda (x)\J \Phi, \left[\J, a_\lambda (y) \right]\Phi \right) \ .
\end{multline}
Again, by Lemma \ref{jaycomrel}, this equals
\begin{equation}
(j_1(x)-1)\Re \left(a_\lambda (x) \J \Phi, \J a_\lambda (y)\Phi \right)
- (j_1(y)-1)\Re
\left(a_\lambda (x)\J \Phi, \J a_\lambda (y)\Phi \right)\ .
\end{equation}
Commuting the $\J$ once more with the $a$'s leads to
\begin{multline}
(j_1(x)-1)\Re \left(\left[a_\lambda (x), \J\right] \Phi,
\J a_\lambda (y)\Phi \right)
- (j_1(y)-1)\Re \left(\left[a_\lambda (x),\J\right] \Phi,
\J a_\lambda (y)\Phi \right) \\
+\left\{j_1(x) -j_1(y)\right\} \Re \left(\J a_\lambda (x)\Phi,
\J a_\lambda (y)\Phi \right) \ .
\end{multline}
We do not have to worry about the last term, since it is the product
of a symmetric and an antisymmetric term in the variables $x$ and $y$,
and thus its $x$-$y$ integral with $|x-y|^{-4}$ in (7) vanishes. The
other term, using Lemma \ref{jaycomrel}, is of the form
\begin{equation}
\left\{(j_1(x)-1)^2 - (j_1(y)-1)(j_1(x)-1)\right\}\Re \left(\J a_\lambda
(x)\Phi,
\J a_\lambda (y)\Phi \right)  \ .
\end{equation}
Taking into account the terms with $x$ and $y$ exchanged we find that
(\ref{remainderjunk}) equals
\begin{equation}\label{fieldcommutator}
4 \pi \sum_\lambda \int \frac{\left(j_1(x)-j_1(y) \right)^2 \Re
\left(\J a_\lambda (x)\Phi, \J a_\lambda (y)\Phi \right) }{
|x-y|^4} d x d y \ .
\end{equation}

Now we use Lemma \ref{outerphotons} to get an estimate on the size of
(\ref{fieldcommutator}).  Recall that the function $j_1(x)$, which
defines $\J$ and which is defined in (\ref{j1j2}), is identically
equal to $1$ on $\cup_{1=1}^nQ_i$, where $Q_i$ is the ball of radius
$L/2$ equi-centered with the ball $B_i$. Moreover $j_1(x) =0$ whenever
the distance of $x$ to the center of every $B_i$ exceeds $L$. Write
$1= \chi_1+\chi_2+\chi_3$ where $\chi_1$ is the characteristic
function of $\cup_{1=1}^n Q_i$ and $\chi_2$ is the characteristic
function of the shell between $\cup_{1=1}^n Q_i$ and $\Sigma =
\cup_{1=1}^n P_i$.  Finally, $\chi_3$ is the characteristic function
of the outside region (on which $j_2(x)=1$ and $j_1(x)=0$). We note,
for later use, that $\int \chi_1 \leq CnL^3$ and $\int \chi_2 \leq
CnL^3$, where $C$ is a universal constant.

Next, we analyze each of the terms
\begin{equation}
T_{i,j}=
4 \pi\int \frac{\left(j_1(x)-j_1(y) \right)^2
\chi_i(x) \chi_j(y) \Re \left(\J a_\lambda (x)\Phi, \J a_\lambda (y)\Phi
\right) }
{|x-y|^4} d x d y \ .
\end{equation}
Clearly, $T_{1,1} =T_{3,3} = 0$.
To bound the other $T_{i,j}$'s  we recall that
\begin{equation}
(j_1(x)-j_1(y))^2 \leq \frac{C }{ L^2} |x-y|^2 \ .
\end{equation}
With this and Schwarz's inequality, $T_{i,j}$  is bounded by
\begin{multline}
  T_{i,j} \leq \frac{C }{ L^2} \int \frac{\chi_i(x) \Vert \J a_\lambda
    (x)\Phi \Vert \
    \chi_j(y)\Vert \J a_\lambda (y)\Phi \Vert }{ |x-y|^2} d x d y \\
  \leq \frac{C }{ L^2} \int \frac{\chi_i(x) \Vert a_\lambda (x)\Phi
    \Vert \ \chi_j(y)\Vert a_\lambda (y)\Phi \Vert }{ |x-y|^2} d x d y
  .
\end{multline}
Denote $\Vert a_\lambda (x)\Phi \Vert $ by $f(x)$.  Consider the terms $i=3$
and $j=1,2$.  Using the Hardy-Littlewood-Sobolev inequality (see
\cite[Theorem 4.3]{Analysis}  with $1/2+5/6+2/3=2$), we get the bound
\begin{equation}
T_{3,j} \leq
\frac{C }{ L^2} \Vert \chi_3 f \Vert_{2} \Vert \chi_j f \Vert_{6/5}\ ,
\ \ \ j=1,2 \ .
\end{equation}
By H\"older's inequality
\begin{equation}
\Vert \chi_j f \Vert_{6/5} \leq \Vert \chi_j \Vert_{3}
\Vert f \Vert_2 \leq Cn^{1/3}L \Vert f \Vert_2 \ ,
\end{equation}
and hence, for $j=1,2$,
\begin{equation}
T_{3,j} \leq
\frac{C n^{1/3}}{ L} (\Phi, {\mathcal{N}} \Phi)^{1/2}
(\Phi, {\mathcal{N}}_{{\rm out}}  \Phi)^{1/2} \ .
\end{equation}
Note that $\Vert \chi_3 f \Vert_{2}$ is proportional to
$(\Phi, {\mathcal{N}}_{{\rm out}}  \Phi)^{1/2}$
while $\Vert f \Vert_{2}$ is proportional to
$(\Phi, {\mathcal{N}} \Phi)^{1/2}$.

The next term to consider is $i=2$ and $j=1,2,3$. Again, H-L-S leads to the
bound
\begin{equation}
T_{2,j} \leq
\frac{C }{ L^2} \Vert \chi_j f \Vert_{2} \Vert \chi_2 f \Vert_{6/5}\ ,
\ j=1,2,3 \ .
\end{equation}
Applying H\"older's inequality yields
\begin{equation}
\Vert \chi_2 f \Vert_{6/5} \leq \Vert \chi_2  \Vert_3 \ \Vert \chi_2 f
\Vert_2 \leq Cn^{1/3}L  \Vert \chi_2 f \Vert_2 \ ,
\end{equation}
and hence the term with $i=2$ and $j=1,2,3$ is also bounded above by
\begin{equation}
\frac{Cn^{1/3} }{ L} (\Phi,  {\mathcal{N}}\Phi)^{1/2}
(\Phi, {\mathcal{N}}_{{\rm out}}
\Phi)^{1/2} \ ,
\end{equation}
where we used $ (\Phi, {\mathcal{N}}_{{\rm out}} \Phi)^{1/2} \geq
\Vert \chi_2 f \Vert_2$. It is the term $(\Phi, \mathcal{N} \Phi)$ which
yields
the logarithmic term in formula (\ref{commutatorexpression}).

Finally, the term $i=1$,\ $j=2$ is the same as $i=2$, \ $j=1$, and the
term $i=1$, \ $j=3$ is the same as $i=3$, \ $j=1$, both of which have
been already treated.

Collecting the estimates we have shown that
$$
\Re \left( \J \Phi, \left[H_f, \J\right] \Phi \right)
\leq \frac{Cn^{1/3} }{ L} (\Phi,  {\mathcal{N}}\Phi)^{1/2}
(\Phi, {\mathcal{N}}_{{\rm out}} \Phi)^{1/2}
$$
which, by Lemma \ref{outerphotons},  proves the lemma.
\hfill \qedsymbol

We are now ready to prove the estimate stated at the end of the proof
of Theorem \ref{locphot}, i.e.,
\begin{lm} \label{loccomm}
For all $L>2R_0$ we have the estimate
\begin{equation}\label{jaycommutator}
\frac{\left(\J \Phi(n), \left[H^0(n), \J\right] \Phi(n) \right) }
{ \left(\J \Phi(n),\J \Phi(n) \right)}
\leq \frac{C}{(L-2R_0)^{\gamma}} \left(\frac{R_0}{L^\gamma}\right)
(1+|\log(\Lambda R_0)|)
\end{equation}
for all $\gamma <1 $. The constant $C$ depends on $n, \Lambda, \gamma$
but not on $R_0$ and $L$.
\end{lm}

{\it PROOF:} The lemma is a direct consequence of Lemmas
\ref{normalization},
\ref{kincomlem} and \ref{fieldcom}.
\hfill\qedsymbol

\begin{lm}[Bound on the error terms] \label{asquare}
For each fixed $x \in \R^3$,
\begin{equation} (\J\Phi, a (h(x-\cdot))^2 \J \Phi) \leq
C_\gamma \frac{1} {D^{2\gamma}}
\max\left\{(\Phi, \mathcal{N} \Phi), \ 1 \right\},
\end{equation}
for any $\gamma <1$, where $D$ is the distance of $x$ to the support of
$j_1$.
The same estimate holds for $a^* (h(x-\cdot))^2$
and for $a^* (h(x-\cdot))a(h(x-\cdot))$ in place of $a(h(x-\cdot))^2$.
\end{lm}

\medskip
{\it PROOF:} Using Lemma \ref{jaycomrel},
\begin{multline}
(\J\Phi, a (h(x-\cdot))^2 \J \Phi) =
(\J\Phi , \J a (h(x-\cdot)j_1(\cdot))^2\Phi) =\\
( a^*(h(x-\cdot)j_1(\cdot))\J^2 \Phi ,  a (h(x-\cdot)j_1(\cdot))\Phi).
\end{multline}
By Schwarz's inequality this is bounded above by
\begin{equation}\label{astar}
\Vert ( a^*(h(x-\cdot)j_1(\cdot))\J^2 \Phi \Vert \,
\Vert  a (h(x-\cdot)j_1(\cdot))\Phi)\Vert \ .
\end{equation}
Similar to the proof of Lemma \ref{kincomlem}, the second factor in
(\ref{astar})
can be bounded as follows.
\begin{align}
\Vert  a(h(x-\cdot)j_1(\cdot)) \Phi \Vert
&\leq \int j_1(y)\  \Vert h(x - y) a(y) \Phi \Vert d y  \nonumber \\
&\leq \int j_1(y)\ |h(x - y)|  \Vert  a(y) \Phi \Vert d y  \nonumber \\
&\leq \left(\int j_1(y)^2\ |h(x - y)|^2 d y\right)^{1/2}
\left( \int \Vert  a(y) \Phi \Vert^2 d y \right)^{1/2} \ . \label{bound}
\end{align}
The second factor here is $(\Phi, \mathcal{N} \Phi)^{1/2}$
while the first factor can be
estimated, using the fact that the support of
$j_1$ and the point $x$ are a distance $D$ apart, as
\begin{equation}
\frac{1}{D^{\gamma}}\left(\int j_1(y)^2\ |x-y|^{2 \gamma}|h(x - y)|^2
d y\right)^{1/2}
\leq \frac{C}{D^{\gamma}}
\end{equation}
by Lemma \ref{coupling}.

The first factor in (\ref{astar}) can be bounded as follows.
\begin{equation}
\Vert ( a^*(h(x-\cdot)j_1(\cdot))\J^2 \Phi \Vert^2 = (\J^2\Phi,
a (h(x-\cdot)j(\cdot))
a^*(h(x-\cdot)j_1(\cdot)) \J^2 \Phi)
\end{equation}
\begin{equation}\label{commm}
= (\J^2 \Phi, [a(h(x-\cdot)j_1(\cdot)), a^*(h(x-\cdot)j_1(\cdot))] \J^2
\Phi)
+\Vert a(h(x-\cdot)j_1(\cdot)) \J^2 \Phi \Vert ^2 \ .
\end{equation}
The first (commutator) term in (\ref{commm})
equals $\int |h(x-y)|^2 j_1(y)^2 dy\
\Vert \J^2 \Phi\Vert^2$,
which is bounded by $C/D^{2\gamma}$. The second term  was treated in
(\ref{bound}) and is thus bounded by $C/D^{2 \gamma} (\Phi, \mathcal{N}
\Phi)$.
(Note that $(\Phi, \mathcal{N} \Phi) \geq (\J \Phi,
\mathcal{N} \J \Phi)$.)

It is immediate that similar estimates hold when
$aa $ is replaced by $a^* a^*$ or by $a^* a$.
\hfill\qedsymbol

\section{Infrared Bounds}  \label{infraredbounds}

In this section we prove Lemma \ref{outerphotons}. It will follow from
infrared bounds similar to the ones proved in \cite{BFS} and
\cite{GLL}, which have been proved to hold for the electrons bound to
the Coulomb potential.  Those infrared bounds are not sufficient,
however, for the localized wave function of the `free' electrons. The
chief reason for this insufficiency is that we need to know the
dependence of the constants in the infrared bounds on the parameter
$R_0$. Ultimately, the trouble stems from the fact that we do not know
the positions of the $n$ localized electrons, not even remotely. Thus,
a direct application of the estimates in \cite{GLL} would lead to
constants that can grow conceivably as $R_0^2$ or even faster for
large $R_0$.  This problem does not occur for the the bound electrons
since, in that case, the electrons are localized by the Coulomb
potential and the infrared bounds do not depend on the parameter
$R_0$.  The theorem below holds for the localized electrons, as well
as for the bound electrons. The proof for bound electrons is easier
and is omitted.

As in Sect. \ref{commutators} the Dirichlet ground state $\Phi_D(n)$
for the `free' electrons localized in in $\Omega (B_1,\dots, B_n)$ is
denoted simply by $\Phi$. Its energy is $E_D^0(n)$. The ground state
for the bound system with Hamiltonian $H^V(N')$ is denoted by $\Psi$.

\begin{lm}[Infrared bounds] \label{infrared}
The following infrared bounds hold for $\Phi$:
\begin{equation}\label{badinfrared}
\Vert \widehat a_\lambda (k) \Phi \Vert \leq \frac{C}{|k|^{3/2}}
\widehat{\chi}_\Lambda(k) \ ,
\end{equation}
\begin{equation} \label{goodinfrared}
\Vert \widehat a_\lambda(k) \Phi \Vert \leq \frac{CR_0 }
{ |k|^{1/2}}\widehat\chi_{\Lambda}(k) \ .
\end{equation}
The vector $\widehat a_\lambda(k) \Phi$ is a sum of $n$ terms of the form
$e^{-ik\cdot Y_j} \widehat{T}_{j,\lambda}(k)$
where $\widehat{T}_{j,\lambda}(k) $ is
 given by
(\ref{tee}) and satisfies the estimates
\begin{equation}\label{deriveinfrared}
\Vert \nabla_{k} \widehat{T}_{j,\lambda}(k) \Vert \leq \frac{CR_0 }
{ |k|^{1/2}(k_1^2+k_2^2)^{1/2}}\widehat\chi_{\Lambda}(k)  +
\frac{CR_0 }{ |k|^{1/2}}
\nabla_{k}\widehat\chi_{\Lambda}(k)
\end{equation}
The vectors $Y_j$ are defined below.
The constant $C$ depends on $n$, on  the ultraviolet cutoff $\Lambda$ and
it is a monotone decreasing function of $R_0$.
Similar bounds but without the factor $R_0$ hold for the
bound electron ground state $\Psi$.
\end{lm}

{\it PROOF:} We prove first the bound (\ref{goodinfrared}) in detail.
As a first step one performs an operator valued gauge transformation
(see \cite[Eq. 47]{GLL}) by
applying the unitary operator (in which $A$ is the vector potential
(\ref{apot}))
\begin{equation}
U(x)=\exp [-i\sqrt{\alpha} \sum_{j=1}^m \phi_j(x) (x-Y_j)\cdot A(Y_j)]
\end{equation}
to the wave function in each of its variables, i.e.,
\begin{equation}
\Phi \to \widetilde{\Phi}=\prod_{i=1}^n  U(x_i) \Phi =:\mathcal{U} \Phi\ .
\end{equation}
Here, $\phi_j$ is a suitably chosen smooth function of compact support and
the $Y_j$ are suitably chosen vectors.
Note that the factors in the product
commute since $A(x)$ commutes with $A(y)$ for all $x$ and $y$.

Next, we describe the functions $\phi_j$.  Consider the balls $B_i(2R_0)$,
which are concentric with the $B_i$ but with twice the radius, and group
them into  clusters according to whether they overlap or not.
We denote the number of these clusters by $m$. Such a cluster of balls $C_j$
has
a diameter that is bounded above by $4R_0$
times the number of balls in the cluster and hence bounded by
$4nR_0$. Denote by $Y_j$ the center of the cluster $C_j$ which is
defined to be the center of the smallest ball that contains all the balls
$B_i(2R_0)$ belonging to that cluster.  We choose functions $\phi_j$ that
are
smooth , supported in the union of the balls $B_i(2R_0)$ that belong
to the cluster $C_j$ and that are identically one on the union of the balls
$B_i$.  In particular
$\sum_{j=1}^m \phi_j(x) = 1$ for $x \in \cup_{i=1}^n B_i$.

The gauge transformation $\mathcal{U}$ transforms the Hamiltonian
$H^0(n)$ into the Hamiltonian
\begin{align}
\widetilde{H}^0(n) = \mathcal{U} H^0(n) \mathcal{U}^*
= \sum_{i=1}^n &\left[ (p_i + \sqrt{\alpha} \widetilde{A}(x_i))^2
+\frac{g}{2} \sqrt{\alpha}\> \sigma_i \cdot B(x_i)
\right]  \nonumber \\
&+ \alpha\sum_{i < j} \frac{1 }{ |x_i-x_j|} + \widetilde{H}_f
\end{align}
where the new field
$\widetilde{A}(x)=\mathcal{U}A(x)\mathcal{U}^*+\alpha^{-1/2}\mathcal{U}
p \mathcal{U}^*
= A(x)+\alpha^{-1/2}\mathcal{U} p \mathcal{U}^*$ is
given by
\begin{equation}
\widetilde{A}(x)=A(x)- \sum_{j=1}^m \phi_j(x) A(Y_j) - \sum_{j=1}^m
\nabla \phi_j(x) (x-Y_j)\cdot A(Y_j) \ ,
\end{equation}
in which $A(x)$ is still given by (\ref{apot}). The transformed field
energy $\widetilde{H}_f$ is given as in (\ref{eq:fielden}) but with
$\widehat a_\lambda(k)$ replaced by the transformed creation and
destruction operators
\begin{equation}
b_{\lambda}(k, X) = \mathcal{U} \widehat a_\lambda(k) \mathcal{U}^*
=\widehat a_\lambda(k) -i\sqrt{\alpha} w_{\lambda}(k, X) \label{bee}
\end{equation}
with
\begin{equation}\label{doubleu}
w_{\lambda}(k, X) = \sum_{i=1}^n \sum_{j=1}^m
\phi_j(x_i)(x_i-Y_j)\cdot \frac{\varepsilon_{\lambda}(k) }{
|k|^{1/2}}e^{-i k \cdot Y_j}\widehat{\chi}_{\lambda}^{\phantom{B}}(k) \ .
\end{equation}
As before, the letter $X$ denotes the vector $(x_1, \dots, x_n)$.
We note that
\begin{equation}\label{nobee}
\widehat a_\lambda^{\phantom{*}}(k)\Phi=
\mathcal{U}^* \left[\widehat a_\lambda^{\phantom{*}}(k)\widetilde\Phi -
i\sqrt{\alpha} w_\lambda (k,X)\widetilde{\Phi}\right] \ .
\end{equation}

Since $\Phi$ satisfies the Schr\"odinger equation we can apply the standard
pull-through formula \cite{BFS, GLL} and compute
\begin{multline}
\left[\widetilde{H}^0(n)-E^0_D(n)\right] \widehat a_\lambda(k)
\widetilde\Phi =
\left[\widetilde{H}^0(n),\widehat a_\lambda(k)\right] \widetilde\Phi  =  \\
2\widehat{\chi}_\Lambda(k)\sqrt{\alpha}\ |k|^{-1/2}
\varepsilon_{\lambda}(k)\cdot \sum_{i=1}^n
(p_i+\sqrt{\alpha}\
\widetilde{A}(x_i))\left\{\sum_{j=1}^m
(e^{-ik\cdot Y_j}-e^{-ik \cdot x_i})\phi_j(x_i) \right\}
\widetilde{\Phi}       \\
+2\widehat{\chi}_\Lambda(k)\sqrt{\alpha}
\sum_{i=1}^n\sum_{j=1}^m(p_i+\sqrt{\alpha} \
\widetilde{A}(x_i)) \cdot    \nabla \phi_j(x_i)
(x_i-Y_j) \cdot \frac{\varepsilon_{\lambda}(k)}{\sqrt{|k|}}
e^{-ik\cdot Y_j} \widetilde{\Phi} \\ +i\frac{g}{2}
\widehat{\chi}_\Lambda(k)\sqrt{\alpha}\ \frac{k \wedge
\varepsilon_{\lambda}(k)} {\sqrt{|k|}}
\cdot \sum_{i=1}^n \sigma_j e^{-ik\cdot x_i }\widetilde{\Phi}
-|k| b_{\lambda}(k,X) \widetilde{\Phi} \ . \label{mess}
\end{multline}

The term $ \sum_{j=1}^m e^{-ik \cdot x_i} \phi_j(x_i)$ stems from the
commutator of $\widehat a_\lambda(k)$ with $A(x_i)$, which yields a
term proportional to $e^{-i k\cdot x_i}$ without any summation on $j$.
Using, however, the relation $\sum_{j=1}^m\phi_j(x_i)=1$, which is
valid for any $x_i$ in $\cup_{j=1}^n B_j$, we get that $ e^{-ik \cdot
  x_i} = \sum_{j=1}^m e^{-ik \cdot x_i} \phi_j(x_i)$. Note that $x_i$
has to be in $\cup_{j=1}^n B_j$ for, otherwise, $\widetilde{\Phi}$
vanishes.

Likewise, the second term in the above formula appears to have bad infrared
behavior, but in order that $\Phi$ does not vanish it is necessary
that $x_i$ be in one of the balls. But then $\nabla \phi_j(x_i)=0$
since $\phi_j$ is constant, and this term does not contribute.

Since $\widetilde{H}^0(n) - E^0_D(n)$ is nonnegative the operator
$\widetilde{H}^0(n) - E^0_D(n)+|k|$ has a bounded inverse $R(k)$. Thus
(\ref{mess}) leads to the equation
\begin{multline}
\widehat a_\lambda(k)\widetilde{\Phi} =\sqrt{\alpha} \sum_{j=1}^m e^{-i
k\cdot
Y_j} 2\widehat{\chi}_\Lambda(k)\ |k|^{-1/2}
\varepsilon_{\lambda}(k)\cdot R(k) \\ \times\sum_{i=1}^n
(p_i+\sqrt{\alpha}
\widetilde{A}(x_i)) (1-e^{ik \cdot (Y_j-x_i)})\phi_j(x_i)
\widetilde{\Phi}       \\
+i\sqrt{\alpha} \sum_{i=1}^n  \frac{g}{2} \widehat{\chi}_\Lambda(k) \frac{k
\wedge
\varepsilon_{\lambda}(k)} {\sqrt{|k|}}
\cdot R(k) \sigma_i e^{-ik\cdot x_i }\widetilde{\Phi}
+i\sqrt \alpha |k| R(k)w_{\lambda}(k,X) \widetilde{\Phi} \
. \label{messtoo}
\end{multline}

As in \cite{GLL}, simple estimates, using Schwarz's inequality, lead to
\begin{multline}\label{tilde}
\Vert \widehat a_\lambda(k) \widetilde{\Phi} \Vert
\leq 2 \widehat{\chi}_\Lambda(k)\sqrt{\alpha}\  |k|^{1/2} \Vert R(k)
\sum_{i=1}^n (p_i+\sqrt \alpha \widetilde{A}(x_i))^2 R(k) \Vert^{1/2} \\
\times
\left( \sum_{i=1}^n(\sum_{j=1}^m \Vert|x_i-Y_j|\phi_j(x_i)\widetilde{\Phi}
\Vert)^2  \right)^{1/2}      \\
+\frac{|g| }{ 2}\widehat{\chi}_\Lambda(k)\sqrt{ \alpha} \ n |k|^{1/2}
\Vert R(k) \widetilde{\Phi} \Vert
+|k| \Vert R(k) \Vert \Vert w_{\lambda}(k,X) \widetilde{\Phi} \Vert \ .
\end{multline}
Note that the index $j$ in the first summand is determined by the ball to
which the electron $i$ belongs.  Therefore, $|x_i - Y_j| \leq 2n R_0$.

Lemma \ref{ferrolemma} states that
\begin{equation}
\frac{g \sqrt \alpha}{2}\sum_{j=1}^N\sigma_j \cdot B(x_j) +H_f  + C \geq 0
\end{equation}
where
\begin{equation}
C = \frac{1}{8\pi^2}g^2\alpha N^2 \int \widehat{\chi}_\Lambda(k)^2 d k \ .
\end{equation}
Thus we also have that
\begin{equation}
\frac{g \sqrt \alpha}{2}\sum_{j=1}^N\sigma_j \cdot B(x_j)
+\widetilde{H}_f  + C \geq 0 \ ,
\end{equation}
(note that the gauge transformation $\mathcal{U}$ commutes with
$B(x)$) and hence
\begin{align}
\beta :&=
\Vert R(k) \sum_{i=1}^n (p_i+\sqrt \alpha \widetilde{A}(x_i))^2 R(k) \Vert
\nonumber\\
&\leq \Vert R(k)  \left(\sum_{i=1}^n [(p_i+\sqrt \alpha
\widetilde{A}(x_i))^2
+\frac{g}{2} \sigma_i B(x_i)]+\widetilde H_f +C
\right) R(k) \Vert \nonumber \\
&=\Vert R(k) \left(\widetilde{H}^0(n)+C\right) R(k) \Vert
\ .
\end{align}
Thus,  by subtracting and adding $E_D^0-|k|$,
\begin{align}
\beta &\leq \Vert R(k) \left(\widetilde{H}^0(n)-E^0_D(n)+|k| \right) R(k)
\Vert + (|E^0_D(n)-|k||+C) \Vert R(k)^2 \Vert \nonumber  \\
&= \Vert R(k) \Vert + (|E^0_D(n)-|k||+C) \Vert R(k)^2 \Vert
\end{align}
where the constant $C$ depends on $\Lambda $ and $n$.
For the last term in (\ref{tilde}) we have that
\begin{equation}\label{wbound}
\Vert w_{\lambda}(k,X) \widetilde{\Phi} \Vert \leq
 \widehat{\chi}_\Lambda(k)|k|^{-1/2} \sum_{j=1}^m \sum_{i=1}^n
\Vert|x_i-Y_j|\phi_j(x_i)
\widetilde{\Phi}\Vert  \ .
\end{equation}

Since $\Vert R(k) \Vert = 1/|k|$, we have that
$\Vert R(k)\widetilde\Phi \Vert \leq  1/|k|$.

By combining these estimates with (\ref{tilde}) and using
$|x_i-Y_j|\phi_j(x_i) \leq 2nR_0$, we obtain the bound
\begin{equation}
\Vert \widehat a_\lambda(k) \widetilde{\Phi} \Vert \leq CR_0|k|^{-1/2}
\Vert \widetilde{\Phi} \Vert \widehat \chi_\lambda(k)\ ,
\end{equation}
where the constant $C$ depends on $\Lambda, n$ and the energy.  This
estimate carries over to the state $\Phi$ by (\ref{nobee})  since
$w_{\lambda}(k,X)$
applied to $\Phi$ satisfies the same estimate, as we see
in (\ref{wbound}).
Note that the energy $E^0_D(n)$ does depend on $R_0$ but it
is monotone decreasing as a function of $R_0$ (by the variational
principle for Dirichlet boundary conditions) and it is
uniformly bounded below.

Next, we observe in (\ref{messtoo}) that $\widehat a_\lambda(k)
\widetilde{\Phi}$ is a sum of $m$ terms of the form $e^{-ik \cdot Y_j}
\widehat{S}_j(k)$ where
\begin{align}
\widehat{S}_{j,\lambda}(k)
&=\sqrt{\alpha}  2\widehat{\chi}_\Lambda(k)\
|k|^{-1/2}
\varepsilon_{\lambda}(k)\cdot R(k) \nonumber \\
&\qquad\qquad\qquad \qquad \times \sum_{i=1}^n
(p_i+\sqrt{\alpha}
\widetilde{A}(x_i)) (1-e^{ik \cdot (Y_j-x_i)})\phi_j(x_i)
\widetilde{\Phi}  \nonumber\\
&\quad +i\sqrt{\alpha}   \frac{g}{2} \widehat{\chi}_\Lambda(k) \frac{k
\wedge
\varepsilon_{\lambda}(k)} {\sqrt{|k|}}
\cdot R(k)\sum_{i=1}^n \sigma_i \phi_j(x_i) e^{-ik\cdot (x_i -Y_j) }
\widetilde{\Phi} \nonumber \\
&\quad +i \sqrt \alpha \widehat{\chi}_\Lambda(k)|k|^{1/2}
\varepsilon_\lambda
(k)\cdot R(k)\sum_{i=1}^n \phi_j(x_i) (x_i-Y_j) \widetilde{\Phi} \ ,
\label{ess}
\end{align}
where we have used the identity
\begin{equation}
\sum_{i=1}^n \sigma_i e^{-ik\cdot x_i} \widetilde{\Phi}
=\sum_{j=1}^m e^{-ik \cdot Y_j} \sum_{i=1}^n \sigma_i \phi_j(x_i)
e^{-ik \cdot (x_i-Y_j)} \widetilde{\Phi} \ .
\end{equation}
Since by (\ref{bee})
\begin{equation}
\widehat a_\lambda (k)\Phi = \mathcal{U}^*\left[\widehat a_\lambda(k) -i
\sqrt \alpha
w_\lambda (k, X)\right]
\widetilde  \Phi
 \ ,
\end{equation}
we obtain that
\begin{equation}
\widehat a_\lambda(k) \Phi = \sum_{j=1}^n e^{-ik \cdot Y_j}
\widehat{T}_{j,\lambda}(k)
\end{equation}
where
\begin{equation}
\widehat{T}_{j,\lambda}(k) = \mathcal{U}^*
\left[\widehat{S}_{j,\lambda}(k) \widetilde \Phi
-i\sqrt \alpha w_\lambda (k, X)
\right]\widetilde{\Phi} \ . \label{tee}
\end{equation}

Differentiating these expressions with respect to $k$ and proceeding
in the same fashion as in the proof of (\ref{goodinfrared}) yields the
estimate (\ref{deriveinfrared}).

Differentiating the polarization vectors (\ref{polvec}) yields the
factor $\sqrt{k_1^2+k_2^2}$ in the denominator of
(\ref{deriveinfrared}) .  The details of the calculation are the same
as the ones in \cite{GLL} and are omitted.  The bound
(\ref{badinfrared} is considerable easier, since its proof does not
require the gauge transformation.  Otherwise the proof is word for
word as the one above.  Finally, the proof of the infrared bounds for
$\Psi$ is a word for word translation of the one given in \cite{GLL}.
Note that the localization radius does not show up in this calculation
since the electrons are exponentially localized in the vicinity of the
origin.  \hfill\qedsymbol

Finally, we come to the main application of the infrared bounds proved
in this section.

\medskip

{\it PROOF  OF LEMMA \ref{outerphotons}:} Using (\ref{infrared}) we can
write
\begin{equation}
(\Phi, \mathcal{N}_{\rm out} \Phi) =  \int_{j_2(x) >0} \Vert a(x) \Phi
\Vert^2 d x
= \int_{j_2(x) >0} \Vert \sum_{j=1}^nT_{j,\lambda}(x-Y_j)  \Vert^2 d x
\end{equation}
which, by Schwarz's inequality is bounded above by
\begin{equation}
\sqrt{n}\sum_{j=1}^n\int_{j_2(x) >0} \Vert T_{j,\lambda}(x-Y_j)\Vert^2  d x
\ .
\end{equation}
For $x$ in the support of $j_2$ we have that $|x-Y_j| > L$ and hence
\begin{equation}
\int_{j_2(x) >0} \Vert T_{j,\lambda}(x-Y_j)  \Vert^2 d x
\leq \frac{1} {L^{2\gamma}}\int |x|^{2\gamma} \Vert T_{j,\lambda}(x)
\Vert^2 d x \ .
\end{equation}
This last term can be related to the derivative
in $k$ space of the function $\widehat{T}_{j,\lambda}(k)$ by the
formula
\begin{equation}\label{moment}
\int |x|^{2 \gamma}  \Vert T_{j,\lambda}(x) \Vert^2 dx = C_{\gamma}
\int \frac{\left( \nabla_k
\widehat{T}_{j,\lambda}(k) , \nabla_k \widehat{T}_{j,\lambda}(k') \right)}
{|k-k'|^{2\gamma+1} }d k dk'
\end{equation}
where the constant $C_{\gamma}$ is given by
\begin{equation}\label{cgamma}
C_{\gamma} = (4\pi)^{-3/2} \frac{\Gamma(\frac{1+2\gamma}{2})}
{\Gamma (1-\gamma)} \ .
\end{equation}
Indeed, writing $Q_{j,\lambda}(x) = xT_{j,\lambda}(x)$ this formula follows
from
\begin{equation}
\int |x|^{2 \gamma-2}  \Vert Q_{j,\lambda}(x) \Vert^2 dx = C_{\gamma}
\int \frac{\left( \widehat{Q}_{j,\lambda}(k) ,  \widehat{Q}_{j,\lambda}(k')
\right)}
{|k-k'|^{2\gamma+1} }d k dk'
\end{equation}
 \cite[Corollary 5.10]{Analysis} and the fact that
\begin{equation}
\left( \widehat{Q}_{j,\lambda}(k) ,  \widehat{Q}_{j,\lambda}(k') \right)
=\left( \nabla_k
\widehat{T}_{j,\lambda}(k) , \nabla_k \widehat{T}_{j,\lambda}(k') \right) \
.
\end{equation}

Using the bound (\ref{deriveinfrared}) a straightforward calculation
shows that the function $\Vert \nabla_k \widehat{T}_{j,\lambda}(k)
\Vert$ is in $L^p$ for all $p <2$. (The relevant term in
(\ref{deriveinfrared}) is the first term on the right side.)  Using
Schwarz's inequality and the Hardy-Littlewood-Sobolev inequality
\cite[Theorem 4.3]{Analysis} we can therefore bound (\ref{moment}) by
\begin{equation}
C_p \left[\int \Vert \nabla_k \widehat{T}_{j,\lambda}(k) \Vert^p d k
\right] ^{2/p}\leq C R_0^2 \ ,
\end{equation}
with $p=6/(5-2\gamma)$, which is strictly less than $2$ for $\gamma
<1$.
To prove (\ref{photonnumber}) we write for some $0 < H $
\begin{equation}
\left(\Phi, \mathcal{N} \Phi\right) = \sum_\lambda \int _{|k| \leq H}
\Vert \widehat a_\lambda (k) \Phi \Vert^2 d k
+ \sum_\lambda \int _{H \leq |k|} \Vert \widehat a_\lambda (k) \Phi
\Vert^2 d k \ ,
\end{equation}
and, using (\ref{goodinfrared}) and (\ref{badinfrared}) we get
\begin{equation}
 C R_0 \int _{|k| \leq H} \frac{1}{|k|} \widehat{\chi}_\Lambda d k + C
\int _{H \leq |k|}  \frac{1}{|k|^{3}}
 \widehat{\chi}_\Lambda d k \ ,
 \end{equation}
 which, optimized over $H$,  leads to (\ref{photonnumber}).

The proof for the state $\Psi$ is carried out in precisely the
same fashion.
\hfill\qedsymbol

\renewcommand{\thesection}{\Alph{section}}

\section*{Appendix A}
\setcounter{section}{1}
\setcounter{equation}{0}
\setcounter{thm}{0}

\begin{lm}\label{ferrolemma}
On $\wedge^N L^2(\mathbb{R}^3; \mathbb{C}^2) \otimes \mathcal{F}$  we have
that
\begin{equation} \label{ferro}
\frac{g \sqrt \alpha}{2}\sum_{j=1}^N\sigma_j \cdot B(x_j) +H_f  +
\frac{1}{8\pi^2}g^2\alpha N^2 \int \widehat{\chi}_\Lambda(k)^2 d k \geq 0 \
.
\end{equation}
\end{lm}

PROOF:  The magnetic field operator can be written in the form
\begin{equation}
\sum_{\lambda =1}^2 \int [c^*_\lambda(k) \widehat a_\lambda(k)+c_\lambda(k)
\widehat{a}^*_\lambda(k)] dk
\end{equation}
where
\begin{equation}
c^*_\lambda(k)=\frac{g\sqrt \alpha}{4\pi}\widehat{\chi}_\Lambda(k)
\sum_j \frac{ik \wedge \varepsilon_\lambda(k)}{\sqrt{|k|}}\cdot \sigma_j
e^{ik\cdot x_j} \ .
\end{equation}
With this notation we can write the Hamiltonian (\ref{ferro}) as
\begin{align}
&\sum_{\lambda=1}^2 \int |k|\left[\widehat{a}^*_\lambda(k) \otimes I +
\frac{1}{|k|} c^*_\lambda(k)\right]
\left[\widehat a_\lambda(k) \otimes I + \frac{1}{|k|} c_\lambda(k)\right] dk
\nonumber \\
-&\sum_{\lambda=1}^2 \int \frac{1}{|k|} c^*_\lambda(k)c_\lambda(k) dk
\end{align}
Here, $I$ denotes the identity operator on spin space.
The first term is nonnegative and crude estimates on the second yield the
lemma.
\hfill\qedsymbol

\section*{Appendix B}
\setcounter{section}{2}
\setcounter{equation}{0}
\setcounter{thm}{0}

In this section we prove the estimates on the coupling functions
$h^i_\lambda(y)$
which are defined by
\begin{equation}
h_\lambda^i(y)= \frac{1}{2 \pi} \int \frac{1}{\sqrt{|k|}}
\varepsilon^i_\lambda(k)
\chi_\Lambda(k) e^{ik\cdot x} dk \ .
\end{equation}
It is important to choose the polarization vectors carefully in order
that their Fourier transforms (from $k$-space to $y$-space) have nice
decay properties as $|y|$ tends to infinity. We shall express these
decay properties in an integrated form.  The reason for that is that
the decay is not uniform with respect to the direction of the $y$
variable. Recall the definitions (\ref{polvec}).
\begin{lm}[Decay of the coupling functions] \label{coupling}
For any $\gamma < 1$ there is a finite constant $C(\gamma)$ such that
\begin{equation}
\sum_{i=1}^3 \sum_{\lambda=1}^2 \int |y|^{2\gamma} |h^i_\lambda(y)|^2 dy
\leq C(\gamma) \ .
\end{equation}
\end{lm}

PROOF: First we compute the gradient of $h^i_\lambda$ in $k$-space. It
is elementary that
\begin{equation}
|\nabla \frac{1}{\sqrt{|k|}} \varepsilon_\lambda(k)| \leq \frac{C}
{\sqrt{|k|}\sqrt{k_1^2+k_2^2}}
\ ,
\end{equation}
where $C$ is some constant. Because $\chi_\Lambda$ is smooth,
\begin{equation}
\sum_{i=1}^3 \sum_{\lambda=1}^2 \int |\nabla \frac{1}{\sqrt{|k|}}
\varepsilon_\lambda(k)
\chi_\Lambda(k)|^p dk \leq C(p)
\end{equation}
for any $p <2$.
We proceed as in (\ref{moment}), (\ref{cgamma}) and write
\begin{equation}
\sum_{i=1}^3 \sum_{\lambda=1}^2 \int |y|^{2\gamma} |h^i_\lambda(y)|^2 dy
=C_\gamma \sum_{i=1}^3 \sum_{\lambda=1}^2 \int \frac{\nabla \overline{
\widehat{h}^i_\lambda (k)}\cdot \nabla \widehat{h}^i_\lambda (k')}
{|k-k'|^{2\gamma+1}} dk dk' \ .
\end{equation}
Again, by the Hardy-Littlewood-Sobolev inequality \cite[Theorem
4.3]{Analysis} this is bounded by
\begin{equation}
C_p\left[\sum_{i=1}^3 \sum_{\lambda=1}^2 \int |\nabla \frac{1}{\sqrt{|k|}}
\varepsilon_\lambda(k)
\chi_\Lambda(k)|^p dk \right]^{2/p} \leq C_p C(p)^{2/p} \ ,
\end{equation}
where $p=6/(5-2\gamma) < 2$ if $\gamma < 1$. \hfill\qedsymbol

\end{document}